\documentclass[journal,10pt]{IEEEtran}

\usepackage{pgfplots} 
\usepackage{comment}
\usepackage{amsfonts,enumitem}
\usepackage{textcomp}
\usepackage{stackengine}
\stackMath

\def\BibTeX{{\rm B\kern-.05em{\sc i\kern-.025em b}\kern-.08em
		T\kern-.1667em\lower.7ex\hbox{E}\kern-.125emX}}
\usepackage{amsmath,amssymb,multirow}
\usepackage{setspace}
\usepackage{listings}
\usepackage{cite}
\usepackage{stackengine}
\usepackage{acronym}
\usepackage{mathtools}
\usepackage{subfigure}
\usepackage{algorithm,algorithmic}
\usepackage{bm}

\pgfplotsset{
	tick label style={font=\small},
	label style={font=\small},
	legend style={font=\small}
}

\usepackage{enumitem}
\usepackage{mwe} 

\newcommand{\rev}[1]{\textcolor{black}{#1}} 
\newcommand{\revv}[1]{\textcolor{black}{#1}} 

\newcommand\abss[1]{\lvert#1\rvert}

\newcommand\norm[1]{\left\lVert #1\right\rVert}
\newcommand\norms[1]{\big\lVert #1\big\rVert}

\newcommand{\pp}{\boldsymbol{p}}
\newcommand{\hh}{\boldsymbol{h}}

\newcommand{\khat}{\widehat{k}}

\newcommand{\brtext}{{\rm{BR}}}
\newcommand{\rutext}{{\rm{RU}}}
\newcommand{\hhbr}{\hh_{\brtext}}
\newcommand{\hhru}{\hh_{\rutext}}

\newcommand{\mH}{\boldsymbol{H}}
\newcommand{\hhbrbig}{\mH_{\brtext}}
\newcommand{\hhrubig}{\mH_{\rutext}}

\newcommand{\hhrut}{\widetilde{\hh}_{\rutext}}

\newcommand{\alphabr}{\alpha_{\brtext}}
\newcommand{\alpharu}{\alpha_{\rutext}}

\newcommand{\snr}{{\rm{SNR}}}
\newcommand{\nmse}{{\rm{NMSE}}}

\newcommand{\mmtilde}{\widetilde{\mm}}
\newcommand{\mmtildee}[1]{\mmtilde_{#1}}

\newcommand{\yytilde}{\widetilde{\yy}}
\newcommand{\yt}{\widetilde{y}}

\newcommand{\yybrev}{\breve{\yy}}
\newcommand{\bmubrev}{\breve{\bmu}}

\newcommand{\Iset}{\mathcal{I}^{(i)}}

\newcommand{\mm}{\boldsymbol{m}}
\newcommand{\cc}{\boldsymbol{c}}
\newcommand{\yy}{\boldsymbol{y}}
\newcommand{\xx}{\boldsymbol{x}}

\newcommand{\sss}{\bm{s}}

\newcommand{\nn}{\boldsymbol{n}}
\newcommand{\bmu}{\boldsymbol{\mu}}
\newcommand{\zetab}{\boldsymbol{\zeta}}

\newcommand{\tbmu}{\widetilde{\bmu}}

\newcommand{\zetar}{\zeta_{\rm{R}}}
\newcommand{\zetai}{\zeta_{\rm{I}}}

\newcommand{\pfail}{p_{{\rm{fail}}}}

\newcommand{\ber}{{{\rm{Ber}}}}

\newcommand{\nue}{N_{{\rm{UE}}}}
\newcommand{\nbs}{N_{{\rm{BS}}}}

\newcommand{\bbp}{\widebar{\boldsymbol{p}}}
\newcommand{\bb}{\boldsymbol{b}}
\newcommand{\bw}{\boldsymbol{w}}
\newcommand{\bff}{\boldsymbol{f}}

\newcommand{\etab}{\boldsymbol{\eta}}

\newcommand{\pt}{\widetilde{p}}

\newcommand{\boldzero}{{ {\boldsymbol{0}} }}
\newcommand{\boldone}{{ {\boldsymbol{1}} }}
\newcommand{\boldonet}{\boldone^\trp}

\newcommand{\diag}{ \mathrm{diag}  }
\newcommand{\thn}[1]{ {#1^{\rm{th} } } }

\newcommand{\tracesmall}[1]{ {{{\rm{tr}}\left\{ #1 \right\}}}  }

\newcommand{\bA}{\boldsymbol{A}}

\newcommand{\bJ}{\boldsymbol{J}}
\newcommand{\bX}{\boldsymbol{X}}
\newcommand{\bS}{\boldsymbol{S}}
\newcommand{\bB}{\boldsymbol{B}}

\newcommand{\bp}{\boldsymbol{p}}

\newcommand{\quot}[1]{``{#1}''}

\newcommand{\mL}{\mathcal{L}}

\newcommand{\aab}{\boldsymbol{a}}

\newcommand{\aabbs}{\boldsymbol{a}_{\rm{BS}}}
\newcommand{\aabue}{\boldsymbol{a}_{\rm{UE}}}

\newcommand{\Dcal}{ \mathcal{D} }

\newcommand{\etabhat}{ \widehat{\boldsymbol{\eta}} }
\newcommand{\alphab}{\widebar{\alpha}}
\newcommand{\etabbar}{ \widebar{\etab} }
\newcommand{\alphabarre}{ \alphab_{\rm{R}} }
\newcommand{\alphabarim}{ \alphab_{\rm{I}} }

\newcommand{\mcrb}{{\rm{MCRB}}}
\newcommand{\crb}{{\rm{CRB}}}
\newcommand{\lb}{{\rm{LB}}}

\newcommand{\lbpp}{\lb_{\pp}}
\newcommand{\crbperpp}{\crb^{{\rm{Perfect}}}_{\pp}}
\newcommand{\crbknownpp}{\crb^{{\rm{KnownLoc}}}_{\pp}}
\newcommand{\crbricianpp}{\crb^{{\rm{Rician}}}_{\pp}}

\newcommand{\alphare}{\alpha_{\rm{R}}}
\newcommand{\alphaim}{\alpha_{\rm{I}}}

\newcommand{\bet}{\boldsymbol{\eta}}

\newcommand{\alphahat}{\widehat{\alpha}}
\newcommand{\pphat}{\widehat{\pp}}
\newcommand{\mmhat}{\widehat{\mm}}
\newcommand{\cchat}{\widehat{\cc}}
\newcommand{\zetabhat}{\widehat{\zetab}}
\newcommand{\zetahat}{\widehat{\zeta}}

\newcommand{\ppbs}{\pp_{\BS}}
\newcommand{\ppris}{\pp_{\RIS}}

\newcommand{\BS}{\text{BS}}
\newcommand{\RIS}{\text{RIS}}

\newcommand{\complexset}[2]{ \mathbb{C}^{#1 \times #2}  }

\newcommand{\complexsett}{ \mathbb{C}  }
\newcommand{\realset}[2]{ \mathbb{R}^{#1 \times #2}  }
\newcommand{\realsetone}[1]{ \mathbb{R}^{#1}  }

\newcommand{\realp}[1]{ \Re \left\{#1\right\}  }
\newcommand{\imp}[1]{ \Im \left\{#1\right\}  }

\newcommand{\hermit}{\mathsf{H}}
\newcommand{\trp}{\mathsf{T}}

\newcommand{\Gammab}{\bm{\Gamma}}

\newcommand{\gammab}{\bm{\gamma}}

\newcommand{\phib}{\bm{\phi}}
\newcommand{\Phib}{\bm{\Phi}}

\newcommand{\phibt}{\widetilde{\phib}}

\newcommand{\mtCN}{{\mathcal{CN}}}
\newcommand{\Imatrix}{{ \boldsymbol{\mathrm{I}} }}

\makeatletter \renewcommand\d[1]{\ensuremath{%
		\;\mathrm{d}#1\@ifnextchar\d{\!}{}}}
\makeatother

\makeatletter
\newcommand*\rel@kern[1]{\kern#1\dimexpr\macc@kerna}
\newcommand*\widebar[1]{%
	\begingroup
	\def\mathaccent##1##2{%
		\rel@kern{0.8}%
		\overline{\rel@kern{-0.8}\macc@nucleus\rel@kern{0.2}}%
		\rel@kern{-0.2}%
	}%
	\macc@depth\@ne
	\let\math@bgroup\@empty \let\math@egroup\macc@set@skewchar
	\mathsurround\z@ \frozen@everymath{\mathgroup\macc@group\relax}%
	\macc@set@skewchar\relax
	\let\mathaccentV\macc@nested@a
	\macc@nested@a\relax111{#1}%
	\endgroup
}
\makeatother

\usepackage{amsthm}

\theoremstyle{remark}

\newtheoremstyle{mytheoremstyle} 
{\topsep}                    
{\topsep}                    
{\upshape}                   
{.5em}                           
{\itshape}                   
{.}                          
{.5em}                       
{}  

\theoremstyle{mytheoremstyle}

\newtheoremstyle{iremark}
{\topsep}   
{\topsep}   
{\upshape}  
{0.2in}       
{\itshape}  
{.}         
{5pt plus 1pt minus 1pt} 
{\thmname{#1}\thmnumber{ \itshape#2}\thmnote{ (#3)}} 

\theoremstyle{iremark}

\newtheorem{remark}{Remark}

\acrodef{RIS}{reconfigurable intelligent surface}
\acrodef{SNR}{signal-to-noise ratio}
\acrodef{ISAC}{integrated sensing and communication}
\acrodef{ISLAC}{integrated sensing, localization, and communication}
\acrodef{LoS}{line-of-sight}
\acrodef{NLoS}{non-line-of-sight}
\acrodef{AoA}{angle-of-arrival}
\acrodef{AoD}{angle-of-departure}
\acrodef{UE}{user equipment}
\acrodef{NF}{near-field}
\acrodef{FF}{far-field}
\acrodef{BS}{base station}
\acrodef{MCRB}{misspecified Cram\'{e}r-Rao bound}
\acrodef{CRB}{Cram\'{e}r-Rao bound}
\acrodef{LB}{lower bound}
\acrodef{MML}{mismatched maximum likelihood}
\acrodef{ML}{maximum likelihood}
\acrodef{MAP}{maximum a-posteriori}
\acrodef{DL}{downlink}
\acrodef{UL}{uplink}
\acrodef{JLFD}{joint localization and failure diagnosis}
\acrodef{pdf}{probability density function}
\acrodef{cdf}{cumulative distribution function}
\acrodef{LS}{least-squares}
\acrodef{OFDM}{orthogonal frequency-division multiplexing}
\acrodef{FIM}{Fisher Information Matrix}
\acrodef{RMSE}{root mean-squared-error}
\acrodef{NMSE}{normalized mean-squared-error}
\acrodef{SISO}{single-input single-output}
\acrodef{MIMO}{multiple-input multiple-output}

\allowdisplaybreaks 

\pgfplotsset{compat=1.18} 
\begin{document}
	\bstctlcite{IEEEexample:BSTcontrol}

	\title{RIS-aided Localization under Pixel Failures}

	\author{Cuneyd Ozturk,\thanks{C. Ozturk is with  the Department of Electrical and Computer Engineering, Northwestern University, Evanston, IL 60208, USA, E-mail: cuneyd.ozturk@northwestern.edu} \emph{Member, IEEE}, Musa Furkan Keskin, \emph{Member, IEEE},
     \\	Vincenzo Sciancalepore,\thanks{V. Sciancalepore is with NEC Laboratories Europe GmbH, Germany. E-mail: vincenzo.sciancalepore@neclab.eu} \emph{Senior Member, IEEE}, Henk Wymeersch,\thanks{M. F. Keskin and H. Wymeersch are with the Department of Electrical Engineering, Chalmers University of Technology, Sweden, E-mails: \{furkan,henkw\}@chalmers.se} \emph{Senior Member, IEEE}, and Sinan Gezici, \thanks{ S. Gezici is with the Department of Electrical and Electronics Engineering, Bilkent University, Ankara, 06800, Turkey, E-mail:	gezici@ee.bilkent.edu.tr} \emph{Senior Member, IEEE}
     \thanks{This work was supported, in part, by the EU H2020 RISE-6G project under grant 101017011 and by the Swedish Research Council project  2022-03007.}}

	\maketitle
	\begin{abstract}

Reconfigurable intelligent surfaces (RISs) hold great potential as one of the key technological enablers for beyond-5G wireless networks, improving localization and communication performance under line-of-sight (LoS) blockage conditions. However, hardware imperfections might cause RIS elements to become faulty, a problem referred to as \textit{pixel failures}, which can constitute a major showstopper especially for localization. In this paper, we investigate the problem of RIS-aided localization of a user equipment (UE) under LoS blockage in the presence of RIS pixel failures, considering the challenging single-input single-output (SISO) scenario. We first explore the impact of such failures on accuracy through misspecified Cram\'{e}r-Rao bound (MCRB) analysis, which reveals severe performance loss with even a small percentage of pixel failures. To remedy this issue, we develop two strategies for joint localization and failure diagnosis (JLFD) to detect failing pixels while simultaneously locating the UE with high accuracy. The first strategy relies on $\ell_1$-regularization through exploitation of failure sparsity. The second strategy detects the failures one-by-one by solving a multiple hypothesis testing problem at each iteration, successively enhancing localization and diagnosis accuracy. Simulation results show significant performance improvements of the proposed JLFD algorithms over the conventional failure-agnostic benchmark, enabling successful recovery of failure-induced performance degradations.

	\end{abstract}

	\section{Introduction}\label{sec_intro}
	
	\subsection{Background and Motivation}
	\Acp{RIS} are envisaged as a key enabling technology towards 6G to reduce the vulnerability of mmWave and sub-THz systems to signal blockages, providing improved communication rate and coverage \cite{RIS_tutorial_2021,RIS_THz_2021,RIS_commag_2021}. Through their dynamic ability to engineer the propagation environment, \acp{RIS} can be optimized in terms of various performance metrics, such as energy efficiency \cite{RIS_EE_TWC_2019,distRIS_EE_2022} and sum-rate \cite{DRL_RIS_JSAC_2020,RIS_sumrate_2020}. While a great deal of papers has been devoted to \ac{RIS} for communication, especially for overcoming \ac{LoS} blockages \cite{RIS_tutorial_2021,RIS_loc_2021_TWC}, \acp{RIS} enjoy several properties that make them attractive for localization as well \cite{elzanaty2021reconfigurable,wymeersch2020radio}. The large aperture of \acp{RIS} enables high resolution in \ac{AoA} and \ac{AoD} estimation, while their functioning over large bandwidths supports high delay resolution (in addition to high data rates) \cite{RIS_SPM_2022}. When \acp{UE} are close to the \ac{RIS}, wavefront curvature (also known as \ac{NF}) allows direct relative localization, even when the \ac{LoS} between the \ac{UE} and \ac{BS} is blocked \cite{LOS_NLOS_NearField_2021,rinchi2022compressive,Cuneyd_WCL_RIS_2022}. When the \ac{RIS} has a known location and orientation, this relative location can be transformed into global coordinates, effectively rendering the \ac{RIS} into an additional analog \ac{BS} \cite{RIS_SPM_2022}. 

 	\begin{figure}[t]
	    \centering
	    \includegraphics[width=0.9\columnwidth]{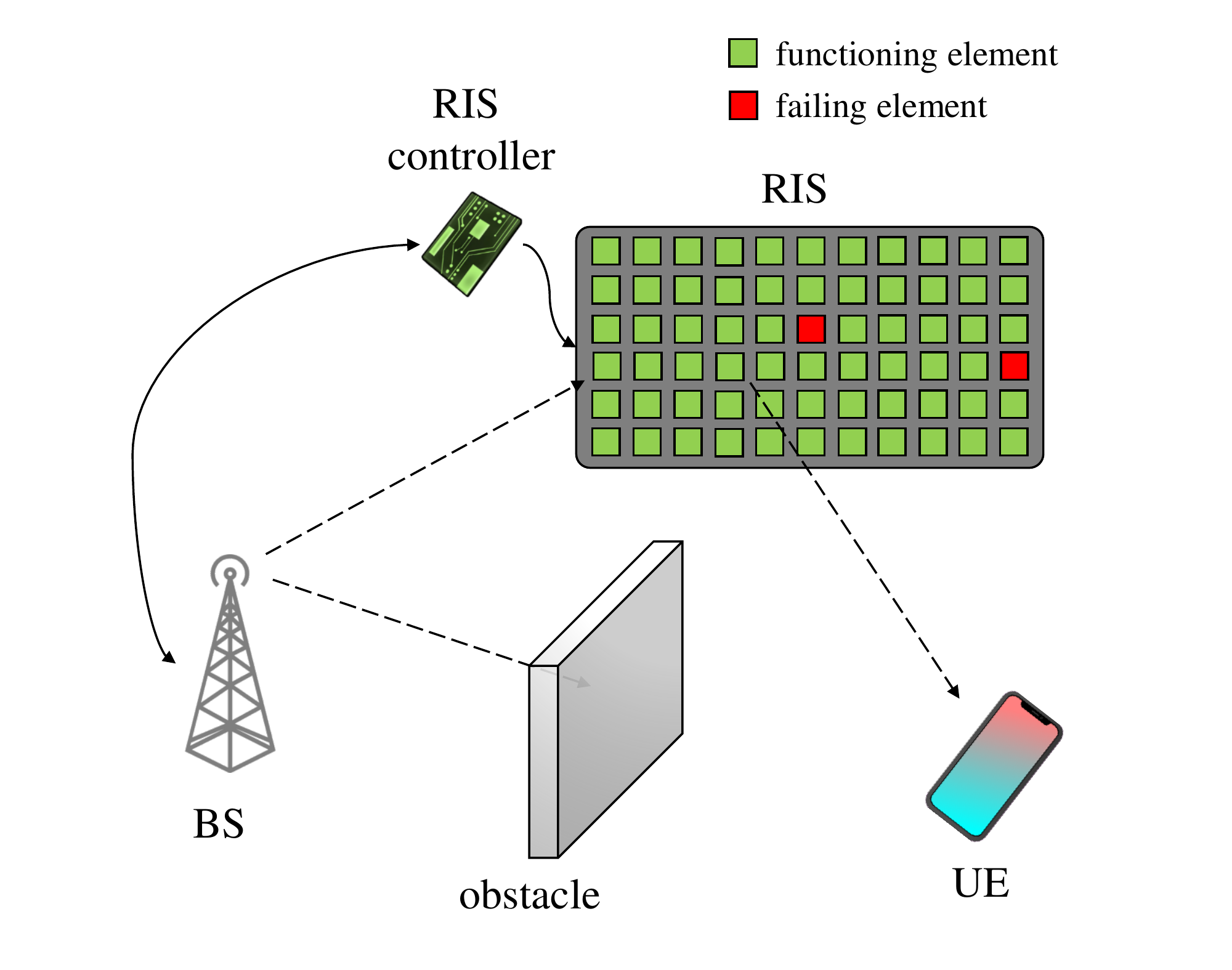}
        \vspace{-0.5cm}
	    \caption{\rev{Scenario under consideration, comprising a single-antenna \ac{BS}, a single-antenna \ac{UE}, and an \ac{RIS}. The \ac{LoS} path between the \ac{BS} and the \ac{UE} is blocked and the downlink communications is achieved through the \ac{RIS}. The colors on the \ac{RIS} elements indicate the status of different pixels (green means properly functioning element, while red indicates failing pixel). The goal is to localize the \ac{UE} in 3D under RIS pixel failures using a set of scalar baseband observations, which leads to the problem of joint localization and RIS diagnosis.}}
	    \label{fig:setup}
	\end{figure}
	
	In light of the above observations, a number of studies has recently investigated \ac{RIS} for \ac{NF} localization \cite{Shaban2021,LOS_NLOS_NearField_2021,cuneyd_ICC_RIS_2022,Cuneyd_WCL_RIS_2022,rinchi2022compressive,rahal2022constrained,luan2021phase,HRIS_NF_2022}. In \cite{Shaban2021}, performance bounds and a practical method for narrowband \ac{NF} localization of a transmitter with an \ac{RIS} acting as a lens has been presented. The studies in \cite{cuneyd_ICC_RIS_2022,Cuneyd_WCL_RIS_2022} consider a more conventional setup including a reflective \ac{RIS} and a \ac{BS} to evaluate \cite{cuneyd_ICC_RIS_2022} and mitigate \cite{Cuneyd_WCL_RIS_2022} the impact of phase-dependent amplitude variations of the \ac{RIS} elements on \ac{NF} localization accuracy. Taking into account the scatterers in the links from \ac{BS} to \ac{RIS} and from \ac{RIS} to \ac{UE}, \cite{rinchi2022compressive} employs compressive sampling and optimized \ac{RIS} configurations to achieve high accuracy. Moreover, \ac{RIS} phase profile optimization for localization in \ac{NF} has been proposed in \cite{rahal2022constrained} based on a combination of positional and derivative beams, which reveals considerable improvements over standard positional beams. The work in \cite{luan2021phase} tackles a similar problem for a linear \ac{RIS}. Furthermore, direct and two-step estimators for \ac{LoS}/\ac{NLoS} \ac{NF} localization using \ac{OFDM} transmission have been developed in \cite{LOS_NLOS_NearField_2021} for a stripe-like \ac{RIS}, where sub-cm level accuracy has been demonstrated. Finally, the study in \cite{HRIS_NF_2022} derives the theoretical limits on \ac{NF} localization with hybrid \ac{RIS} and determines localization-optimal RIS configurations. 

	\subsection{Related Work on Failures}
	An important consideration in a practical \ac{RIS}, which may comprise several hundreds of unit elements (or \emph{pixels}), is that individual elements may fail. This problem also exists in the array processing literature. 
	In \cite{mmwave_array_diagnosis_2018}, antenna array diagnosis has been studied in a standard mmWave setup without a \ac{RIS} and several compressive sensing based techniques have been proposed to identify the faulty antenna elements and the resulting amplitude and phase distortions. The \ac{AoA} estimation problem under element failures has been considered in \cite{ULA_DOA_Sensor_Failure}, where the diagnosis of faulty elements is formulated as a Toeplitz matrix reconstruction problem. Recently, several papers investigate RIS element failures in mmWave communications \cite{errorAnalysis_RIS_failure_2020,RIS_diagnosis_2021}. In \cite{errorAnalysis_RIS_failure_2020}, the authors present different types of pixel errors (e.g., stuck at state, out of state, etc.) and their spatial distribution (independent, clustered, etc.). The paper also explores the effect of pixel errors on the radiation pattern through simulation analysis. The study in \cite{RIS_diagnosis_2021} establishes a failure model to specify the amplitude and phase shift of faulty elements and proposes diagnostic methods by exploiting the sparsity property of failures. In \cite{phaseCalib_RIS_TSP_2022}, joint phase error calibration and channel estimation algorithm has been proposed to deal with RIS phase offsets induced by hardware impairments. Moreover, \cite{Doppler_RIS_HWI_2021} evaluates the impact of failures and phase quantization errors on the \ac{SNR} of \ac{RIS}-aided high-mobility vehicular communications. 
 
 In addition to element failures resulting from internal hardware imperfections, RIS element blockages due to external environmental effects such as dust, rain and ice have been studied in recent works \cite{Joint_RIS_BS_Diagnosis_2021,Diagnosis_CE_RIS_2020}. In \cite{Joint_RIS_BS_Diagnosis_2021}, blockages at both the BS and RIS are considered and an iterative algorithm is proposed to jointly estimate the blockage coefficients of the BS and RIS. In \cite{Diagnosis_CE_RIS_2020}, the authors propose a two-stage algorithm for joint RIS diagnosis and channel estimation in a RIS-aided mmWave \ac{MIMO} system. Despite a considerable amount of research on pixel failures in \textit{RIS-aided communications}, to the best of the authors' knowledge, no study has been performed to tackle the problem of \textit{RIS-aided localization} in the presence of pixel failures. Hence, two fundamental questions arise that remain unanswered so far: \textit{(i)} how severe can the impact of \ac{RIS} element failures be on the localization accuracy?, and \textit{(ii)} is it possible to perform \ac{RIS} diagnosis and \ac{UE} localization simultaneously?


\subsection{Contributions}\label{sec_contr}
In this paper, we address both of these knowledge gaps, and perform an in-depth analysis of RIS pixel failures in terms of achievable localization accuracy. We also propose novel algorithms to identify faulty pixels and mitigate their impact on localization, even when the \ac{UE} location is a-priori unknown. 
The main contributions can be listed as follows:
\begin{itemize}
    \item \textbf{Localization under RIS Pixel Failures:} For the first time, we investigate the problem of RIS-aided localization under pixel failures by adopting a practical failure model based on biases in individual failing elements \cite{errorAnalysis_RIS_failure_2020,RIS_diagnosis_2021}. We consider a narrowband geometric near-field scenario where a single-antenna \ac{UE} estimates its location using downlink signals by a single-antenna \ac{BS} in the presence of a \ac{RIS} with \ac{LoS} blockage. The biases can change the resulting RIS phase profiles and lead to a challenging problem of \ac{JLFD}.
    
    \item \textbf{Impact of Failures on Localization Accuracy:} We carry out a theoretical analysis on how detrimental RIS pixel failures can be for RIS-aided localization. To quantify the effect of failures on the localization performance, we \rev{utilize} the \ac{MCRB} \cite{Fortunati2017} \rev{as a theoretical tool to serve as a fundamental benchmark} under mismatch between a true model with pixel failures and an assumed model without failures (i.e., the \ac{UE} is unaware of failures). \rev{This analysis} reveals fundamental insights on when such failures can induce severe degradations in localization accuracy (i.e., SNR regimes, number of failing elements, etc.).
    
	\item \textbf{Sparsity-Inspired Joint Localization and Failure Diagnosis Algorithms:} \rev{We derive the hybrid \ac{ML}/\ac{MAP} estimator} for the \ac{JLFD} problem, which involves joint estimation of the \ac{UE} location and the locations and failure coefficients (i.e., biases) of the faulty elements. \rev{To cope with the combinatorial nature of the problem,} we propose two \rev{novel \ac{JLFD}} algorithms \rev{that} exploit the sparsity of failures \rev{to provide computationally feasible solutions}. The first one, called $\ell_1$-\ac{JLFD}, is based on an $\ell_1$-regularization approach that estimates the failure mask and the \ac{UE} location using an alternating optimization strategy. The second algorithm, called Successive-\ac{JLFD}, explicitly considers the statistics of pixel failures and solves \rev{the optimization problem} in an iterative manner, in a spirit similar to that of OMP based sparse channel estimation algorithms \cite{OMP_mmWave_2016,mmWave_pos_TWC_2018,grossi2020adaptive}. In particular, Successive-\ac{JLFD} detects the failures one-by-one at each iteration and proceeds by canceling out their impact on the subsequent iterations, progressively improving the performance of localization and failure mask estimation. 
\end{itemize}

Additionally, simulation results showcase the sensitivity of localization performance to pixel failures, which indicates the need for powerful methods to counteract such impairments. In the challenging \ac{SISO} \ac{RIS} scenario under consideration, the proposed \ac{JLFD} algorithms are shown to recover performance losses resulting from pixel failures and achieve accuracy levels very close to theoretical limits corresponding to the case with known failure mask (i.e., perfectly calibrated/diagnosed RIS).

\rev{\textit{Notations:} $\odot$ denotes the Hadamard (element-wise) product. $\diag(\xx)$ outputs a diagonal matrix with the elements of a vector $\xx$ on the diagonals. $[\xx]_n$ denotes the $\thn{n}$ entry of a vector $\xx$, while $[\bX]_{:,n}$ represents the $\thn{n}$ column of a matrix $\bX$. An all-ones vector of appropriate size is indicated by $\boldone$.}
 

\vspace{-0.1in}
\section{System Model and Problem Description}\label{sec:System}
In this section, we describe the system geometry and the signal model, introduce the RIS pixel failure model, and formulate the problem of localization under pixel failures.
\vspace{-0.1in}
    
\subsection{Geometry and Signal Model}\label{sec_sig_model}
\vspace{-0.05in}
Consider an RIS-aided \ac{DL} localization scenario consisting of a single-antenna BS, an $N$-element RIS, and a single-antenna UE, as shown in Fig.~\ref{fig:setup}. The BS is located at a known position $\ppbs \in \realsetone{3}$, while $\ppris \in \realsetone{3}$ denotes the known center of the RIS and $\pp_{n} \in \realsetone{3}$ represents the known location of the $n$-th RIS element for $1\leq n\leq N$. The UE has an unknown location $\pp \in \realsetone{3}$, which needs to be estimated. 
	
For \ac{DL} communications, the BS transmits narrowband pilots $\sss = [ s_1 \ \cdots \ s_T ]^\trp \in \complexset{T}{1}$ over $T$ transmission instances under an average power constraint $\mathbb{E}\left\{\abss{s_t}^2\right\} = E_s$. To motivate the deployment of a \ac{RIS}, we assume that the \ac{LoS} path between the BS and the UE is blocked \cite{RIS_tutorial_2021,RIS_loc_2021_TWC}, e.g., due to buildings, cars or trees. \rev{The \ac{RIS} acts as a passive reflector with controllable phase shifts to relay the pilots from the BS to the UE.} In addition, it is assumed that no uncontrolled multipath (i.e., those paths induced by scattering/reflection off the passive objects in the environment) exists\footnote{\label{fn_mult}The effect of uncontrolled multipath can be removed from the received signal via temporal coding of RIS phase profiles \cite{LOS_NLOS_NearField_2021,RIS_SISO_JSTSP_2022}. \rev{Temporal coding can be applied by setting $\phib_{2t-1} = \phibt_t, ~ \phib_{2t} = -\phibt_t$ and forming a new set observations $\yt_t = (y_{2t-1} - y_{2t})/2$ for $t = 1, \ldots, T/2$, where $y_t$ and $\phib_t$ will be defined in \eqref{eq_yt} and \eqref{eq_gammab}, respectively.}}. Then, the \ac{DL} communications occur only through the \ac{RIS} and the \ac{DL} signal received by the UE at transmission $t$ is given by   
\begin{equation}\label{eq_yt}
	y_t = \alpha \, \aab^\trp(\pp) \diag(\gammab_t) \aab(\ppbs) s_t + n_t\,, 
\end{equation}
where $\alpha$ is the unknown channel gain, $\gammab_t \in \mathbb{C}^{N\times 1}$ is the RIS phase profile at transmission $t$, and $n_t$ is zero-mean additive white Gaussian noise with variance $N_0/2$ per real dimension. Moreover, $\aab(\pp) \in \complexset{N}{1}$ denotes the \ac{NF} RIS steering vector for a given position $\pp$, which can be expressed by taking the RIS center $\ppris$ to be the reference point as \cite{EM_wavefront}, \cite{nearfieldTrack_TSP_2021}
\begin{align} \label{eq_nf_steer}
	[\aab(\pp)]_{n} = \exp\left(-j \frac{2\pi}{\lambda}\left(\norm{\pp-\pp_n}-\norm{\pp-\ppris}\right)\right)
\end{align}
for\rev{\footnote{\label{fn_nf_siso}\rev{Wavefront curvature manifested in the \ac{NF} steering vector \eqref{eq_nf_steer} through location-dependent phase shifts enables \ac{SISO} localization from the narrowband observations in \eqref{eq_yt}. Note that the spherical wavefront model in \eqref{eq_nf_steer} is generic and covers both the \ac{NF} and \ac{FF} cases. Hence, the performance analysis in Sec.~\ref{sec:Problem1} and the proposed methods in Sec.~\ref{sec:JointLocFailDiag} are valid even when the UE is located beyond the Fraunhofer distance (i.e., usually considered as the \ac{FF} region \cite{Fresnel_2011,Fresnel_2016,nearfieldTrack_TSP_2021}) $d = \norm{\pp-\ppris} \geq 2 D^2/\lambda$, where $D$ is the RIS aperture size (i.e., the largest distance between any two RIS elements). This will be verified through simulations in Sec.~\ref{sec_perf_dist}.}}} $n\in\{1, \ldots, N\}$, \rev{where $\lambda$ denotes the carrier wavelength}. For convenience, let us define $\yy = [y_1 \, \cdots \, y_T]^\trp$, $\bb(\pp) = \aab(\pp) \odot \aab(\ppbs)$, $\Gammab = [\gammab_1 \, \cdots \, \gammab_T] \in \complexset{N}{T}$, $\bS = \diag(s_1, \ldots, s_T)$ and $\nn = [n_1 \, \cdots \, n_T]^\trp$. Then, the observations in \eqref{eq_yt} can be written compactly as
\begin{align}\label{eq_y_vec}
	\yy = \alpha \bS \Gammab^\trp \bb(\pp) + \nn ~,
\end{align}
where $\nn \sim \mtCN(\boldzero, \allowbreak N_0 \Imatrix ) $.

\vspace{-0.15in}
\rev{\begin{remark}[Extension to MIMO Scenarios]\label{remark_mimo}
    The \ac{SISO} signal model in \eqref{eq_yt} can be readily extended to MIMO scenarios as \revv{(assuming analog arrays)} \cite{RIS_tutorial_2021}
\begin{equation}\label{eq_yt_r1}
	y_t = \alpha \, \bw^\trp  \hhrubig \diag(\gammab_t)  \hhbrbig \bff s_t + n_t\,,
\end{equation}
where $\bw \in \complexset{\nue}{1}$ and $\bff \in \complexset{\nbs}{1}$ denote, respectively, the \revv{analog} combiner at the UE and the \revv{analog} precoder at the BS, and $\nue$ and $\nbs$ are the array sizes at the UE and the BS, respectively. In addition, 
\begin{align} \label{eq_hhbrbig_m}
    \hhbrbig &= \aab(\ppbs) \aabbs^\trp(\ppris-\ppbs) \in \complexset{N}{\nbs} \,, \\ \label{eq_hhrubig_m}
    \hhrubig &= \aabue(\ppris-\pp) \aab^\trp(\pp) \in \complexset{\nue}{N} \,
\end{align}
represent the BS-RIS and the RIS-UE channel matrices, where $\aabbs(\cdot) \in \complexset{\nbs}{1}$ and $\aabue(\cdot) \in \complexset{\nue}{1}$ denote the array steering vectors of the BS and the UE, respectively, as a function of displacement with respect to a given position. Defining $\widetilde{\alpha} \triangleq \alpha \, \bw^\trp  \aabue(\ppris-\pp) \aabbs^\trp(\ppris-\ppbs) \bff$ and inserting \eqref{eq_hhbrbig_m} and \eqref{eq_hhrubig_m}, we can recast \eqref{eq_yt_r1} as
\begin{equation}\label{eq_yt_siso2_m}
	y_t = \widetilde{\alpha} \,  \aab^\trp(\pp) \diag(\gammab_t)  \aab(\ppbs)  s_t + n_t\,,
\end{equation}
which is equivalent to the SISO signal model in \eqref{eq_yt}. Hence, the performance analysis in Sec.~\ref{sec:Problem1} and the proposed JLFD algorithms in Sec.~\ref{sec:JointLocFailDiag} can be applied to MIMO scenarios, as well.
\end{remark}}

\vspace{-0.15in}
\subsection{RIS Pixel Failure Model}\label{sec_failure_model}
To model RIS pixel/element failures, we consider \textit{biases} in individual RIS elements, where the element switches to a valid, biased state with a certain distance from the desired state due to bit-flipping or external biases \cite{errorAnalysis_RIS_failure_2020,RIS_diagnosis_2021}. Under such element failures, the RIS phase profile $\gammab_t$ in \eqref{eq_yt}  can be modeled as
\begin{align}\label{eq_gammab}
	\gammab_t = \phib_t \odot \mm \,,
\end{align}
where $\phib_t \in \complexset{N}{1}$ represents the \textit{configurable} RIS weights under the designer's control \rev{(known to the entity performing localization)}, and $\mm = [m_1 \, \cdots \, m_N] \in \complexset{N}{1}$ denotes the unknown \textit{failure mask} quantifying the effect of faulty elements, which can be defined as \cite{mmwave_array_diagnosis_2018,RIS_diagnosis_2021}
	\begin{align}\label{eq_mask_m}
	m_n &= \begin{cases}  \zeta_n, & {\textrm{if the $\thn{n}$ RIS element is faulty (biased)}} \\
		1, & \textrm{if the $\thn{n}$ RIS element is functioning}
	\end{cases} .
\end{align}
In \eqref{eq_mask_m}, $\zeta_n = \kappa_n e^{j \psi_n}$ denotes the failure related complex response of the $n$-th element, with $ 0 < \kappa_n \leq 1$ and $0 \leq \psi_n < 2 \pi$ representing the resulting attenuation and phase shift, respectively. 

We assume a stochastic failure model where each RIS element fails independently from each other with the probability $\pfail$. In addition, when the $\thn{n}$ pixel fails, its complex response $\zeta_n$ follows the distribution $\kappa_n\sim\mathcal{U}(0,1)$ and $\psi_n\sim\mathcal{U}(-\pi,\pi)$ \cite{RIS_diagnosis_2021}. Formally, each element $m_n$ of the failure mask in \eqref{eq_mask_m} can be expressed as \cite{CS_AMP_2013}
\begin{align}\label{eq_mn}
    m_n = c_n \zeta_n + 1-c_n ~,
\end{align}
where the binary variable $c_n \in \{0,1\}$ specifies the absence/presence of failure, with $c_n \sim \ber(\pfail)$, and $\zeta_n \in \complexsett$ corresponds to the complex amplitude of the failing element in case of failure. According to \eqref{eq_mn}, when $c_n = 0$ (i.e., no failure), we have $m_n = 1$, while $c_n = 1$ (i.e., failure) sets $m_n = \zeta_n$, in compliance with \eqref{eq_mask_m}. Hence, $m_n$ has a \textit{spike-and-slab} prior \cite{CS_AMP_2013,spike_slab_JMLR_2014,spike_slab_PAMI_2013}, given by
\begin{align}\label{eq_pdf_mn}
    f_{m_n}(m_n) = (1-\pfail) \delta(m_n-1) + \pfail f_{\zeta_n}(m_n) ~,
\end{align}
where $m_n$ attains the \quot{spike} value $1$ if the $\thn{n}$ element is functioning and is drawn from the \quot{slab} \ac{pdf} $f_{\zeta_n}(m)$ in case of failure (see App.~\ref{app_pdf} for the derivation of $f_{\zeta_n}(m)$).  


Defining $\Phib = [\phib_1 \, \cdots \, \phib_T] \in \complexset{N}{T}$, the RIS phase profiles $\Gammab$ in \eqref{eq_y_vec} can be expressed under the failure model \eqref{eq_gammab} as
\begin{align} \label{eq_gammab_profiles}
	\Gammab = \Phib \odot \mm \boldonet ~,
\end{align}
where the \ac{pdf} of the failure mask can be written using \eqref{eq_pdf_mn} and under the assumption of independently failing elements as
\begin{align}\label{eq_mm_pdf}
    f_{\mm}(\mm) = \prod_{n=1}^{N} f_{m_n}(m_n) ~.
\end{align}
With pixel failures in \eqref{eq_gammab_profiles}, the observation model \eqref{eq_y_vec} becomes
\begin{align}\label{eq_y_vec_fail}
	\yy = \alpha \bS \left(\Phib^\trp \odot \boldone \mm^\trp \right) \bb(\pp) + \nn ~.
\end{align}

\vspace{-0.2in}
\subsection{Problem Description for Joint Localization and Failure Diagnosis}\label{sec_prob_desc}
Given the observations $\yy$ in \eqref{eq_y_vec_fail} and the prior distribution of the failure mask $\mm$ in \eqref{eq_mm_pdf}, the problem of joint localization and RIS failure diagnosis\rev{\footnote{\label{fn_jlfd}\rev{ In principle, all pixel failures can be detected and calibrated before deploying the RIS. However, during normal operation, pixel failures can occur at any time and the UE may be situated in any arbitrary location, which requires detecting these failures dynamically while performing localization. Consequently, the problems of failure detection and UE localization are intertwined, giving rise to the \ac{JLFD} problem under consideration. As will be shown in Sec.~\ref{sec_mcrb_numerical} and Sec.~\ref{sec_perf_alg}, localization ignoring failures experiences severe degradations in accuracy.}}} is to estimate\rev{\footnote{\label{fn_est_jlfd}\rev{\ac{JLFD} can be performed either at the UE or at the BS, depending on the computational capability of the UE. For instance, when the computational resources of the UE are limited, it can send its measurements \eqref{eq_yt} to the BS over the uplink via the RIS, enabling the BS to perform \ac{JLFD}. }}} the UE position $\pp$, the channel gain $\alpha$ and the failure mask $\mm$. To tackle this problem, we first characterize lower bounds on localization accuracy in the presence of pixel failures in Sec.~\ref{sec:Problem1}, with the aim to evaluate performance losses due to such impairments. In Sec.~\ref{sec:JointLocFailDiag}, we formulate the problem in a mathematically rigorous manner and propose two algorithms to solve it, followed by their complexity analysis in Sec.~\ref{sec:CompAnalysis}.


\section{Localization Performance Evaluation under Pixel Failures}\label{sec:Problem1}
In this section, we derive theoretical limits on localization in the presence of pixel failures under varying levels of knowledge regarding the failing pixels. To this end, we resort to the \ac{MCRB} \cite{Fortunati2017} as a tool to assess degradation in localization performance due to mismatch between the ideal/assumed model with no failures and the true model with pixel failures. 
In addition, we employ standard \ac{CRB}, as well, to evaluate theoretical performance under perfect knowledge of failing pixel locations and perfect/imperfect knowledge of associated complex coefficients.

\subsection{\ac{MCRB} Analysis under Pixel Failures}\label{sec_mcrb}
In this part, we quantify localization performance for the case where the UE is unaware of pixel failures and therefore estimates its location by assuming that all pixels are functioning (i.e., $\mm = \boldone$ in \eqref{eq_y_vec_fail}). We leverage the \ac{MCRB} analysis
to characterize theoretical limits on localization accuracy under the aforementioned conditions \cite{Fortunati2017,Cuneyd_WCL_RIS_2022}. 

\subsubsection{True and Assumed Models}
We first describe the true and assumed models in the presence of pixel failures. 
\paragraph{True Model}  
According to the \ac{MCRB} terminology \cite{Fortunati2017, Fortunati2018Chapter4}, the true model is given by \eqref{eq_y_vec_fail}, i.e.,
\begin{align}\label{eq_y_vec_fail_true}
	\yy = \alphab \bS \left(\Phib^\trp \odot \boldone \mm^\trp \right) \bb(\bbp) + \nn ~,
\end{align}
where $\alphab$ and $\bbp$ denote the true values of the unknown parameters $\alpha$ and $\pp$, respectively. For a given failure mask $\mm$, the \ac{pdf} of the true model in \eqref{eq_y_vec_fail_true} can be written as
\begin{align}\label{eq_p_true}
    p(\yy) = \frac{1}{(\pi N_0)^{T} } \exp\left\{ - \frac{ \norm{\yy - \bmu }^2 }{N_0} \right\} ~,
\end{align}
where $\bmu \triangleq \alphab \bS \left(\Phib^\trp \odot \boldone \mm^\trp \right) \bb(\bbp) \in \complexset{T}{1}$.

\paragraph{Assumed Model}
For the assumed model, we consider an ideal \ac{RIS} without pixel failures (i.e., $\mm = \boldone$ in \eqref{eq_y_vec_fail}), leading to
\begin{align}\label{eq_y_vec_fail_assumed}
	\yy = \alpha \bS \Phib^\trp \bb(\pp) + \nn ~,
\end{align}
in which case the misspecified parametric \ac{pdf} is obtained as \cite{Fortunati2017}
\begin{align}\label{eq_p_assumed}
    \pt(\yy  \lvert \bet) = \frac{1}{(\pi N_0)^{T} } \exp\left\{ - \frac{ \norm{\yy - \tbmu(\bet) }^2 }{N_0} \right\} ~,
\end{align}
where $\bet = \left[\alphare , \alphaim, \bp^\trp\right]^\trp$ represents the unknown parameters, $\alphare \triangleq \realp{\alpha}$, $\alphaim \triangleq \imp{\alpha}$ and $\tbmu(\bet) \triangleq \alpha \bS \Phib^\trp \bb(\pp) \in \complexset{T}{1}$.

\subsubsection{Pseudo-True Parameter}
The pseudo-true parameter is used in the \ac{MCRB} derivation and is defined as \cite{Fortunati2017}
\begin{equation}\label{eq_eta0}
		\etab_0 = \arg \min_{\etab} ~ \Dcal \left( p(\yy) ~ \Vert ~ \pt(\yy \lvert \bet) \right)~,
\end{equation}
which minimizes the Kullback-Leibler (KL) divergence $\Dcal \left( p(\yy) ~ \Vert ~ \pt(\yy \lvert \bet) \right)$ between the true and assumed \acp{pdf}. In \eqref{eq_eta0}, $\etab_0$ can be found using \cite[Lemma~1]{Cuneyd_WCL_RIS_2022}.

\subsubsection{MCRB and LB}
The covariance matrix of any misspecified-unbiased (MS-unbiased) estimator of $\etab_0$ can be lower-bounded by the \ac{MCRB} matrix \cite{Fortunati2017}:
\begin{align}\label{eq_mcrb_mat}
		\mathbb{E}_p\{(\etabhat(\yy)-\etab_0)(\etabhat(\yy)-\etab_0)^\trp\} \succeq  \mcrb(\etab_0) ~,
\end{align}
where $\mathbb{E}_p\left\{\cdot\right\}$ represents the expectation over the true pdf in \eqref{eq_p_true} and $\etabhat(\yy)$ denotes an MS-unbiased estimator of $\etab_0$ based on the misspecified model \eqref{eq_y_vec_fail_assumed}, meaning that $\mathbb{E}_p\left\{ \etabhat(\yy) \right\} = \etab_0$. Hereafter, $\etabhat(\yy)$ will be referred to as the \textit{failure-agnostic estimator} and employed as a benchmark in performance evaluations in Sec.~\ref{sec:NumRes}. The \ac{MCRB} matrix in \eqref{eq_mcrb_mat} is given by
\begin{align} \label{eq_mcrb_def}
		\mcrb(\bet_0) = \bA_{\etab_0}^{-1} \bB_{\bet_0} \bA_{\bet_0}^{-1} \in \realset{5}{5} ~,
\end{align}
where $\bA_{\etab_0} \in \realset{5}{5}$ and $\bB_{\etab_0} \in \realset{5}{5}$ are defined as \cite{Fortunati2017,Cuneyd_WCL_RIS_2022}
\begin{align}\label{eq_Aeta0}
		[\bA_{\bet_0}]_{i,j} &= \mathbb{E}_p\left\{\frac{\partial^2}{\partial \eta_i \partial \eta_j} \log \pt(\yy  \lvert \bet) \Big|_{\bet = \bet_0}\right\}, 
		\\ \label{eq_Beta0}
		[\bB_{\bet_0}]_{i,j} &= \mathbb{E}_p\left\{\frac{\partial}{\partial \eta_i } \log \pt(\yy  \lvert \bet) \frac{\partial}{\partial \eta_j } \log \pt(\yy  \lvert \bet) \Big|_{\bet =  \bet_0}\right\}.
\end{align} 
Using \eqref{eq_mcrb_def}, the covariance matrix of $\etabhat(\yy)$ with respect to the true value $\etabbar \triangleq [\alphabarre, \alphabarim, \bbp^\trp]^\trp$ satisfies
\begin{align} \label{eq_lb_ms_unb}
		\mathbb{E}_p\{(\etabhat(\yy)-\etabbar)(\etabhat(\yy)-\etabbar)^T\} \succeq  \lb(\bet_0)~,
\end{align}
where the \ac{LB} is obtained as
\begin{align}\label{eq_lb}
	  \lb(\bet_0) = \mcrb(\bet_0)  + (\etabbar-\bet_0)(\etabbar-\bet_0)^\trp~.
\end{align}
From \eqref{eq_lb}, the theoretical lower bound on the localization accuracy under pixel failures is given by
\begin{align} \label{eq_lb_pp}
    \lbpp = \tracesmall{[\lb(\bet_0)]_{3:5,3:5}} ~.
\end{align}

\subsection{Standard \ac{CRB} Analysis under Pixel Failures}
We carry out standard \ac{CRB} analysis to characterize theoretical performance when the UE is aware of pixel failures, considering varying levels of knowledge on the failure mask. Our goal is to evaluate the gap between the \ac{MCRB}-based lower bound in \eqref{eq_lb_pp} and the standard \ac{CRB}, which reveals the degree of performance loss when pixel failures are ignored.

\subsubsection{CRB-Perfect}
In this case, the \ac{UE} has perfect knowledge of $\mm$ in \eqref{eq_y_vec_fail}. The corresponding \ac{FIM} $\bJ(\bet_1)\in\mathbb{R}^{5\times 5}$ for $\bet_1 = \left[\alphare , \alphaim, \bp^\trp\right]^\trp$ in \eqref{eq_y_vec_fail} can be expressed as \cite[Eq.~(15.52)]{kay1993fundamentals} 
	\begin{equation}\label{eq:FIM}
	    \bJ(\bet_1) = \frac{2}{N_0} \realp{ \left(\frac{\partial \bmubrev(\bet_1)}{\partial \bet_1}\right)^\hermit \left(\frac{\partial \bmubrev(\bet_1)}{\partial \bet_1}\right) },
	\end{equation}
	where $\bmubrev(\bet_1) \triangleq \alpha \bS \left(\Phib^\trp \odot \boldone \mm^\trp \right) \bb(\pp) \in \complexset{T}{1}$ and $\frac{\partial \bmubrev(\bet)}{\partial \bet_1} \in \complexset{T}{5}$. The \ac{CRB} on UE location estimation is then computed as
	\begin{align}\label{eq_crb_perf}
    \crbperpp = \tracesmall{[\bJ^{-1}(\bet_1)]_{3:5,3:5} } ~.
\end{align}

	 
\subsubsection{CRB-KnownLoc}
 This case corresponds to the CRB when the locations of failing elements are known, but the respective failure coefficients ($\zeta_n$'s in \eqref{eq_mask_m}) are taken as unknowns. Specifically, the unknown parameter vector is given by
 	\begin{equation}
	   \bet_2 = \left[\alphare , \alphaim, \bp^\trp,  \{\kappa_n\}_{n\in\mathcal{I}} , \{\theta_n\}_{n\in\mathcal{I}}\right]^\trp ~,
	\end{equation}
where $\mathcal{I}$ denotes the set of failure indices, i.e., $m_n = 1$ for $n \notin \mathcal{I}$ and $m_n \neq 1$ for $n \in \mathcal{I}$. The corresponding \ac{FIM} $\bJ(\bet_2)\in\mathbb{R}^{(5+2\lvert\mathcal{I}\rvert)\times (5+2\lvert\mathcal{I}\rvert)}$ can be computed using \begin{equation}\label{eq:FIM2}
	    \bJ(\bet_2) = \frac{2}{N_0} \realp{ \left(\frac{\partial \bmubrev(\bet_2)}{\partial \bet_2}\right)^\hermit \left(\frac{\partial \bmubrev(\bet_2)}{\partial \bet_2}\right) },
	\end{equation}
where $\bmubrev(\bet_2)$ is as defined in \eqref{eq:FIM} and $\frac{\partial \bmubrev(\bet_2)}{\partial \bet_2} \in \complexset{T}{(5+2\lvert\mathcal{I}\rvert)}$. From \eqref{eq:FIM2}, the CRB on localization can be obtained as
\begin{align}\label{eq_crb_known}
    \crbknownpp = \tracesmall{ [\bJ^{-1}(\bet_2)]_{3:5,3:5} } ~.
\end{align}
	
	

\section{Joint Localization and Failure Diagnosis}\label{sec:JointLocFailDiag}
In this section,  we rigorously formulate the \ac{JLFD} problem described in Sec.~\ref{sec_prob_desc} via a hybrid ML/MAP estimation approach \cite{Hybrid_ML_MAP_TSP}, considering the existence of both deterministic (UE position and channel gain) and random (failure mask) unknown parameters. Due to the NP-hardness of the resulting mixed-integer problem, we develop two algorithms to solve its certain approximated versions by exploiting RIS failure sparsity.

\subsection{Hybrid ML/MAP Estimator}\label{sec_hyb_ml_map}
Based on the prior distribution of $\mm$ in \eqref{eq_mm_pdf}, the hybrid ML/MAP estimator for the \ac{JLFD} problem formulated in Sec.~\ref{sec_prob_desc} can be written as \cite{Hybrid_ML_MAP_TSP}
\begin{align}\label{eq_hybrid_ml_map}
    (\alphahat, \pphat, \mmhat  ) = \arg \max_{\alpha, \pp, \mm} f_{\yy, \mm} (\yy, \mm; \alpha, \pp) 
\end{align}
where the goal is to estimate the deterministic position $\pp$ and gain $\alpha$, and the random failure mask $\mm$. In \eqref{eq_hybrid_ml_map},
\begin{align}\label{eq_pdf_all}
    f_{\yy, \mm} (\yy, \mm; \alpha, \pp) = f_{\yy \lvert \mm} (\yy \lvert \mm; \alpha, \pp) f_{\mm}(\mm) ~,
\end{align}
represents the joint \ac{pdf} of $\yy$ and $\mm$, $f_{\yy \lvert \mm} (\yy \lvert \mm; \alpha, \pp)$ is the conditional \ac{pdf} of $\yy$ given $\mm$, and $f_{\mm}(\mm) $ is the prior \ac{pdf} of $\mm$ in \eqref{eq_mm_pdf}. From \eqref{eq_mm_pdf} and \eqref{eq_y_vec_fail}, the log-likelihood in \eqref{eq_pdf_all} can be expressed as
\begin{align} \nonumber
    \log f_{\yy, \mm} (\yy, \mm; \alpha, \pp) &\propto - \frac{\norm{	\yy - \alpha \bS \left(\Phib^\trp \odot \boldone \mm^\trp \right)  \bb(\pp) }^2}{N_0}
    \\ \label{eq_log_fym}
    &~~+ \sum_{n=1}^{N} \log f_{m_n}(m_n) ~.
\end{align}

According to \eqref{eq_mn} and \eqref{eq_pdf_mn}, optimization over $\mm = [m_1 \, \cdots \, m_N] \in \complexset{N}{1}$, with each element having a spike-and-slab prior \ac{pdf}, should be performed by jointly estimating the binary failure vector $\cc = [c_1 \, \cdots \, c_N]^\trp \in \{0, 1\}^N$ (i.e., spikes) and the failure amplitudes vector $\zetab = [\zeta_1 \, \cdots \, \zeta_N]^\trp \in \complexset{N}{1}$ (i.e., slabs). Hence, \eqref{eq_hybrid_ml_map} can be recast using \eqref{eq_pdf_all} and \eqref{eq_log_fym} as
\begin{subequations} \label{eq_hybrid_ml_map_mip}
\begin{align}\nonumber
    (\alphahat, \pphat, \cchat, \zetabhat  ) = \arg \min_{\alpha, \pp, \cc, \zetab} & \Bigg\{ \frac{\norm{	\yy - \alpha \bS \left(\Phib^\trp \odot \boldone \mm^\trp \right)  \bb(\pp) }^2}{N_0}
    \\ \label{eq_hybrid_ml_map_mip_obj} &~~~~~- \sum_{n=1}^{N} \log f_{m_n}(m_n)
    \Bigg\}
    \\ \label{eq_m_cons} ~~ \mathrm{s.t.} &~~ m_n = c_n \zeta_n + 1-c_n , \, \forall n ~,
    \\ \label{eq_c_cons} & ~~ c_n \in \{0, 1\}, \, \forall n ~,
\end{align}
\end{subequations}
where the prior \ac{pdf} $f_{m_n}(m_n)$ is given in \eqref{eq_pdf_mn}. The problem \eqref{eq_hybrid_ml_map_mip} represents a mixed-integer \rev{non-linear} programming \rev{(MINLP) problem} 
with binary variables $\cc$ and continuous variables $\alpha$, $\pp$ and $\zetab$ \rev{\cite{wolsey2007mixed, Pierre_2012_NPHard, liberti2019undecidability}.} 
We develop two heuristic algorithms to solve it, as described next.





\vspace{-0.2in}
\subsection{$\ell_1$-Regularization Based Joint Localization and Failure Diagnosis}\label{sec_l1_jlfd}
\rev{The core technical challenge in solving \eqref{eq_hybrid_ml_map_mip} pertains to the binary variable $\cc$, which renders the problem computationally intractable.} A possible remedy to circumvent the combinatorial nature of the JLFD problem in \eqref{eq_hybrid_ml_map_mip} is to discard the prior-related term (i.e., the second one) in the objective \eqref{eq_hybrid_ml_map_mip_obj} and estimate $\mm$ directly without estimating $\cc$ and $\zetab$ separately, using only the data-fitting term (i.e., the first one). At first glance, this might seem attractive as $\mm$ appears linearly in the data-fitting term of \eqref{eq_hybrid_ml_map_mip_obj}, which enables closed-form estimation. However, since $T < N$ in practice due to small number of transmissions $T$ and large RIS sizes $N$, the problem of estimating $\mm \in \complexset{N}{1}$ from $\yy \in \complexset{T}{1}$ using only the first term in \eqref{eq_hybrid_ml_map_mip_obj} becomes an under-determined \ac{LS} problem, leading to infinitely many solutions.

To tackle this challenge, we propose to make the sparsity assumption that the number of faulty elements is small compared to the RIS size \cite{RIS_diagnosis_2021}, exploiting the fact that $\pfail$ is usually small in practice\footnote{Such sparsity assumptions have been made in both the mmWave array diagnosis literature \cite{mmwave_array_diagnosis_2018} and the RIS diagnosis studies \cite{RIS_diagnosis_2021}. In a \ac{RIS}-aided scenario, the sparsity assumption can be readily justified by noting that for each observation period consisting of $T$ transmissions as in \eqref{eq_y_vec}, the \ac{UE} always detects and estimates additional failures that occur during the latest observation window by having already calibrated the previous ones in the previous periods.}. Under this sparsity assumption, we propose an $\ell_1$-regularization based \ac{JLFD} method, called $\ell_1$-JLFD hereafter, where estimates of $\mm$, $\pp$ and $\alpha$ are updated in an alternating manner, as detailed in the following.

\subsubsection{Update $\mm$ for fixed $\alpha$ and $\pp$ via $\ell_1$-regularization}
For a given $\alpha$ and $\pp$, we formulate the problem of failure mask recovery in \eqref{eq_hybrid_ml_map_mip} as an $\ell_1$-regularized \ac{LS} problem (i.e., the LASSO problem \cite{tibshirani2013lasso})
\begin{align}\label{eq_l_1_m_est_v1}
   \mmhat = \arg \min_{\mm} &  \norm{	\yy - \alpha \bS \left(\Phib^\trp \odot \boldone \mm^\trp \right)  \bb(\pp) }_2^2 + \xi \norm{\mm-\boldone}_1 ~,
\end{align}
where $\xi$ denotes the regularization parameter that governs the trade-off between data-fitting and sparsity. Since most of the elements of $\mm$ are $1$ due to small $\pfail$, $\mm-\boldone$ will be a sparse vector, which is enforced in \eqref{eq_l_1_m_est_v1} via $\ell_1$-regularization. The problem \eqref{eq_l_1_m_est_v1} can be recast in a more convenient LASSO form as
\begin{align}\label{eq_l_1_m_est}
   \mmhat = \arg \min_{\mm} &  \norm{	\yy - \bA(\alpha, \pp) \mm  }_2^2 + \xi \norm{\mm-\boldone}_1 ~,
\end{align}
where
\begin{align} \label{eq_up}
    \bA(\alpha, \pp) \triangleq \alpha \bS \left(\Phib^\trp  \odot \boldone \bb^\trp(\pp) \right) \in \complexset{T}{N} ~.
\end{align}
The problem \eqref{eq_l_1_m_est} can be solved using off-the-shelf convex solvers \cite{cvx} or some standard methods, such as iterative shrinkage/thresholding algorithm (ISTA) \cite{beck2009fast}.

\subsubsection{Update $\alpha$ and $\pp$ for fixed $\mm$}
For a given failure mask $\mm$ in \eqref{eq_hybrid_ml_map_mip}, the problem of estimating $\alpha$ and $\pp$ can be written as
\begin{align}\label{eq_pos_given_m}
   (\alphahat, \pphat) = \arg \min_{\alpha, \pp} &  \norm{	\yy - \alpha \bS \left(\Phib^\trp \odot \boldone \mm^\trp \right)  \bb(\pp) }_2^2  ~,
\end{align}
which can be solved via \cite[Alg.~1]{Cuneyd_WCL_RIS_2022}.

The overall $\ell_1$-JLFD algorithm, which alternates between updating $\mm$ via \eqref{eq_l_1_m_est} and updating $\alpha$ and $\pp$ via \eqref{eq_pos_given_m}, is provided in Algorithm~\ref{alg_ell_1}.

\begin{algorithm}[t]
	\caption{$\ell_1$-Regularization Based Joint Localization and Failure Diagnosis ($\ell_1$-JLFD) Algorithm to Solve \eqref{eq_hybrid_ml_map_mip}}
	\label{alg_ell_1}
	\begin{algorithmic}[1]
		\STATE \textbf{Input:} Observation $\yy$ in \eqref{eq_y_vec_fail}, RIS phase profiles $\Phib$, convergence threshold $\varepsilon$ and maximum number of iterations $M$.
		\STATE \textbf{Output:} UE location $\pphat$, failure mask $\widehat{\mm}$ and channel gain $\alphahat$.
		\STATE \textbf{Initialization:} Set $i=0$. Initialize the failure mask to be the all-ones vector, i.e., $\mm^{(0)} = \boldone$. Compute the corresponding $\pp^{(0)}$ and $\alpha^{(0)}$ via \cite[Alg.~1]{Cuneyd_WCL_RIS_2022}.
		\STATE \textbf{Iterations:}
		\WHILE {$i < M$}
		\STATE Given $\alpha^{(i)}$ and $\pp^{(i)}$, estimate $\mm^{(i+1)}$ by solving the LASSO problem in \eqref{eq_l_1_m_est}. 
		\STATE Given $\mm^{(i+1)}$, estimate $\pp^{(i+1)}$  and $\alpha^{(i+1)}$ in \eqref{eq_pos_given_m} via \cite[Alg.~1]{Cuneyd_WCL_RIS_2022}.
		\IF {($\norm{\pp^{(i+1)}-\pp^{(i)}}_2\leq \varepsilon$) or $i = M$}
		\STATE Set $\widehat{\mm} = \mm^{(i)}$.
		\STATE \textbf{break}
		\ENDIF
		\STATE Set $i = i + 1$.
		\ENDWHILE
		\STATE \textbf{Refinement:} Refine the estimates of the UE location and channel gain for the final failure mask $\mmhat$ via \cite[Alg.~1]{Cuneyd_WCL_RIS_2022}. 
	\end{algorithmic}
	\normalsize
\end{algorithm}

\subsection{Successive Joint Localization and Failure Diagnosis}\label{sec_jlfd}
The $\ell_1$-JLFD algorithm considered in Sec.~\ref{sec_l1_jlfd} provides a convenient way of tackling the NP-hard JLFD problem \eqref{eq_hybrid_ml_map_mip}; however, it does not fully exploit the statistical characteristics of pixel failures specified in \eqref{eq_pdf_mn}--\eqref{eq_mm_pdf}. In this part, we develop a successive failure detection and mask estimation algorithm that detects the pixel failures one-by-one per iteration and estimates the corresponding failure coefficients \rev{by heuristically solving \eqref{eq_hybrid_ml_map_mip} in an iterative manner. This approach} progressively improves mask estimation and localization performance over iterations. The developed algorithm effectively exploits the prior distribution of pixel failures in \eqref{eq_pdf_mn} and is similar in spirit to orthogonal matching pursuit (OMP) \cite{mp_93} type sparse channel estimation/sensing algorithms that extract paths/targets one-by-one, e.g., \cite{OMP_mmWave_2016,mmWave_pos_TWC_2018,grossi2020adaptive}. In particular, at each iteration, we detect the pixel that is most likely to fail assuming at most one failure, based on the posterior probabilities of corresponding pixel failure events given the observation $\yy$ (quantified through the cost function \eqref{eq_hybrid_ml_map_mip_obj} of the hybrid ML/MAP estimator), similar to detection of the strongest path/target in OMP. The details of the proposed algorithm to solve \eqref{eq_hybrid_ml_map_mip} are provided below.


\subsubsection{Initialization}\label{sec_init_alg2}
We begin by computing initial estimates $\pp^{(0)}$, $\alpha^{(0)}$ of position and channel gain. To this end, we assume no pixel failures occur (i.e., we set the initial mask estimate as $\mm^{(0)} = \boldone$) and use \cite[Alg.~1]{Cuneyd_WCL_RIS_2022} \rev{as a small subroutine} to initialize position and channel gain \rev{given a fixed mask}. 

\subsubsection{First Iteration}\label{sec_first_iter}
Given the initial position and channel gain estimates $\pp^{(0)}$, $\alpha^{(0)}$, in the first iteration we assume that at most one pixel fails and formulate a multiple hypothesis testing problem involving $N+1$ different hypotheses corresponding to individual failures of $N$ pixels and the no-failure case. That is, 
 \begin{align} \label{eq_hypo}
	\mathcal{H}_0&: \text{no failure}, \\ \nonumber
	\mathcal{H}_k&:  \thn{k} \text{ pixel fails for } k = 1, \ldots, N.
\end{align}
Stated more formally, under the assumption of \textit{at most single pixel failure} in \eqref{eq_hypo}, the problem \eqref{eq_hybrid_ml_map_mip} branches into $N+1$ subproblems, where the constraint \eqref{eq_c_cons} of the $\thn{k}$ subproblem is given by $c_k = 1, c_n = 0 ~ \forall n \neq k$ for $k = 1, \ldots, N$ and  $c_n = 0 ~ \forall n$ for $k = 0$. This implies that according to the constraint \eqref{eq_m_cons}, the mask for the $\thn{k}$ subproblem, denoted by $\mmtildee{k}$, is given by   
\begin{subequations}\label{eq_mtildek}
\begin{align} 
\mmtildee{0} &= \boldone ~, \\ 
	[\mmtildee{k}]_n&= \begin{cases}  1, & \textrm{if } n\neq k\\ \label{eq_mtildekn}
		\zeta_k, & \textrm{if $n = k$}
	\end{cases},~ k = 1, \ldots, N ~.
\end{align}
\end{subequations}
With the given initial estimates $\alphahat = \alpha^{(0)}$ and $\pphat = \pp^{(0)}$, and $\mmtildee{k}$ defined in \eqref{eq_mtildek}, the cost function associated to the $\thn{k}$ subproblem of \eqref{eq_hybrid_ml_map_mip} can then be formulated as
\begin{align}\nonumber
    \mL(\mmtildee{k}) =&  \frac{\norm{	\yy - \alpha^{(0)} \bS \left(\Phib^\trp \odot \boldone (\mmtildee{k})^\trp \right)  \bb(\pp^{(0)}) }_2^2}{N_0}
    \\ \label{eq_hybrid_ml_map_mip_k}  &~~~~~- \sum_{n=1}^{N} \log f_{m_n}\big([\mmtildee{k}]_n\big) 
     ~.
\end{align}

We note from \eqref{eq_mtildek} that \eqref{eq_hybrid_ml_map_mip_k} should be minimized over the complex coefficient $\zeta_k$ of the $\thn{k}$ failing pixel for $\mathcal{H}_k$ for $k = 1, \ldots, N $, while it has a fixed value for the no-failure hypothesis $\mathcal{H}_0$. Hence, the $\thn{k}$ subproblem of \eqref{eq_hybrid_ml_map_mip} reads
\begin{align}\label{eq_zetak_general}
    \zetahat_k = \arg \min_{\zeta_k} & ~ \mL(\mmtildee{k})
\end{align}
 for $k = 1, \ldots, N $. Using \eqref{eq_pdf_mn}, the second term in \eqref{eq_hybrid_ml_map_mip_k} can be computed as
\begin{align} \label{eq_logfm}
    &\sum_{n=1}^{N} \log f_{m_n}([\mmtildee{k}]_n) \\ \nonumber
    & = \begin{cases}
    (N-1) \log (1-\pfail) + \log \pfail + \log f_{\zeta_k}(\zeta_k), & \textrm{if $k \geq 1$}
    \\
    N \log (1-\pfail), & \textrm{if $k = 0$}
    \end{cases}.
\end{align}
Re-arranging the first term in \eqref{eq_hybrid_ml_map_mip_k}, inserting \eqref{eq_logfm} into the second term and discarding constant terms, the $\thn{k}$ subproblem in \eqref{eq_zetak_general} can be re-written as
\begin{align}\label{eq_zetak}
    \zetahat_k = \arg \min_{\zeta_k} & ~ \frac{ \norms{\yy-\bA(\alpha^{(0)}, \pp^{(0)}) \mmtildee{k} }_2^2}{N_0} + \log \abss{\zeta_k} 
\end{align}
where $\bA(\alpha, \pp)$ is defined in \eqref{eq_up} and the \ac{pdf} of $\zeta_k$ is inserted through \eqref{eq_pdf_zeta}. Since the first (observation-related) term in \eqref{eq_zetak} dominates over the second (prior-related) one at high SNRs, we propose to solve a simpler version of \eqref{eq_zetak}:
\begin{align}\label{eq_zetak2}
    \zetahat_k = \arg \min_{\zeta_k} & ~  \norms{\yytilde_k^{(0)} - \zeta_k \big[\bA(\alpha^{(0)}, \pp^{(0)}) \big]_{:, k}  }_2^2 ~,
\end{align}
where $\yytilde_k^{(0)} \triangleq \yy - \sum_{n=1,n \neq k}^{N} \big[\bA(\alpha^{(0)}, \pp^{(0)}) \big]_{:, n}$. In obtaining \eqref{eq_zetak2} from \eqref{eq_zetak}, we omit the second term in \eqref{eq_zetak} and use \eqref{eq_mtildekn}. The failure coefficient in \eqref{eq_zetak2} can be obtained via \ac{LS} as
\begin{align}\label{eq_zetakhat}
    \zetahat_k = \frac{ \big(\big[\bA(\alpha^{(0)}, \pp^{(0)} \big) \big]_{:, k})^\hermit \yytilde_k^{(0)}}{ \norms{\big[\bA(\alpha^{(0)}, \pp^{(0)})\big]_{:, k})}_2^2 } ~.
\end{align}

Now that we have obtained the failure coefficients $\zeta_k$ in \eqref{eq_zetakhat}  and thus the masks in \eqref{eq_mtildek} for all the hypotheses in \eqref{eq_hypo}, we can compute the corresponding cost functions through \eqref{eq_hybrid_ml_map_mip_k} and select the most likely one:
\begin{align} \label{eq_khat}
	\khat = \arg \min_{0 \leq k \leq N} &~ \mL(\mmtildee{k}) ~,
\end{align}
where $\zeta_k$ is replaced by its estimate in \eqref{eq_zetakhat} to compute $\mmtildee{k}$ in \eqref{eq_mtildek}.
Depending on the outcome of \eqref{eq_khat}, we follow different steps to determine the estimates of position, channel gain, mask and the set of failing pixels, denoted by $\pp^{(1)}$, $\alpha^{(1)}$, $\mm^{(1)}$, $\mathcal{I}^{(1)}$, respectively, at the end of the first iteration:
\begin{itemize}
    \item \textit{No Failure:} If $\khat = 0$, we declare that there is no pixel failure and terminate the algorithm without proceeding to the second iteration\rev{\footnote{\label{fn_alg2_rare}\rev{This implies that in the absence of failures, Algorithm~\ref{alg_successive_complete} will terminate at the first iteration, which only requires calculating the closed-form solution in \eqref{eq_zetakhat} and evaluating the cost function in \eqref{eq_khat} for $N+1$ hypotheses. This feature ensures a computationally cheap and adaptive JLFD solution especially when pixel failures occur rarely.}}}. This yields $\pp^{(1)} = \pp^{(0)}$, $\alpha^{(1)} = \alpha^{(0)}$, $\mm^{(1)} = \mm^{(0)}$ and $\mathcal{I}^{(1)} = \emptyset$.
    
    \item \textit{Failure:} If $\khat \geq 1$, we update the position and channel gain estimates via \cite[Alg.~1]{Cuneyd_WCL_RIS_2022} using the new mask $\mmtildee{\khat}$. For the updated position and channel gain, $\zetahat_k$ and the masks are re-computed via \eqref{eq_zetakhat} and \eqref{eq_mtildekn}, and the hypothesis selection are performed again via \eqref{eq_khat}. We perform these alternating steps until the number of allowed steps is exceeded or the change in the position estimates becomes negligible (typically, this takes $2-3$ steps), yielding the selected hypothesis $\khat$ and the corresponding estimates $\pp^{(1)}$, $\alpha^{(1)}$ and $\mm^{(1)} = \mmtildee{\khat}$ in the end. In this case, we set $\mathcal{I}^{(1)} = \{\khat\}$.
\end{itemize}

\subsubsection{$\thn{i}$ Iteration}
At the $\thn{i}$ iteration ($i \geq 2$), we perform similar operations as in the first iteration with certain changes. Specifically, the number of hypotheses reduces to $N+1-\left\lvert\mathcal{I}^{(i-1)}\right\rvert$, i.e.,
 \begin{align} \label{eq_hypo_i}
	\mathcal{H}_0&: \text{no additional failure}, \\ \nonumber
	\mathcal{H}_k&:  \thn{k} \text{ pixel fails for } k \in \{1, \ldots, N\} \setminus \mathcal{I}^{(i-1)} ~,
\end{align}
where $ \mathcal{I}^{(i-1)}$ represents the estimated pixel failure index set at the end of the $(i-1)$th iteration.

Hence, given the failure mask $\mm^{(i-1)}$ and the position and gain estimates $\pp^{(i-1)}$ and $\alpha^{(i-1)}$, we tackle the \ac{JLFD} problem \eqref{eq_hybrid_ml_map_mip} at the $\thn{i}$ iteration under the assumption of at most one additional failure to decide on the most likely hypothesis in \eqref{eq_hypo_i}. To this end, we define the masks corresponding to the hypotheses in \eqref{eq_hypo_i}, similar to \eqref{eq_mtildek}, as
\begin{subequations}\label{eq_mtildek_i}
\begin{align} 
\mmtildee{0} &= \mm^{(i-1)} ~, \\ \label{eq_mtildek_i2}
	[\mmtildee{k}]_n &= \begin{cases}  1, & \textrm{if } n\neq k ~ \textrm{and}~ n\not\in\mathcal{I}^{(i-1)}
 \\ 
  [\mm^{(i-1)}]_{n}, & \textrm{if}~ n\neq k ~ \textrm{and}~ n\in\mathcal{I}^{(i-1)}
 \\ 
		\zeta_k, & \textrm{if $n = k$}
	\end{cases} ~.
\end{align}
\end{subequations}
Using \eqref{eq_up}, the cost function in \eqref{eq_hybrid_ml_map_mip_k} should then be modified as
\begin{align} \label{eq_hybrid_ml_map_mip_k_i}
    \mL(\mmtildee{k}) =&  \frac{\norm{	\yy - \bA(\alpha^{(i-1)}, \pp^{(i-1)}) \mmtildee{k}  }_2^2}{N_0}
    - \sum_{n=1}^{N} \log f_{m_n}\big([\mmtildee{k}]_n\big) 
     ~,
\end{align}
where the second term is given in App.~\ref{app_comp_2nd}.

Following a similar line of reasoning as in the first iteration, the complex coefficient of the hypothesized failing element for $\mathcal{H}_k$, $k\geq 1$, can be computed via
\begin{align}\label{eq_zetak2_i}
    \zetahat_k = \arg \min_{\zeta_k} & ~  \norms{\yytilde_k^{(i-1)} - \zeta_k \big[\bA(\alpha^{(i-1)}, \pp^{(i-1)}) \big]_{:, k}  }_2^2 ~,
\end{align}
where $\yytilde_k^{(i-1)} \triangleq \yy - \sum_{n=1,n \neq k}^{N} [\mmtildee{k}]_n \big[\bA(\alpha^{(i-1)}, \pp^{(i-1)}) \big]_{:, n}$. Similarly to \eqref{eq_khat}, the most likely hypothesis can be decided via
\begin{align} \label{eq_khat2}
	\khat = \arg \min_{k \in \{0, \ldots, N\} \setminus \mathcal{I}^{(i-1)}} &~ \mL(\mmtildee{k}) ~,
\end{align}
where $\mL(\cdot) $ is defined in \eqref{eq_hybrid_ml_map_mip_k_i} and $\mmtildee{k}$ is given by \eqref{eq_mtildek_i2} with $\zeta_k$ replaced by $\zetahat_k$ in \eqref{eq_zetak2_i}.

Based on \eqref{eq_khat2}, two paths can be followed:
\begin{itemize}
    \item \textit{No Additional Failure:} If $\khat = 0$, no additional pixel failure is detected at the $\thn{i}$ iteration and we terminate the algorithm with the current values of position, channel gain and failure mask.

    \item \textit{Additional Failure:} If $\khat \geq 1$, we detect a failure at the pixel location $\khat$ in addition to the existing ones in $\mathcal{I}^{(i-1)}$. In this case, given the new set of failing pixels $\mathcal{I}^{(i)} = \mathcal{I}^{(i-1)}  \cup \{\khat\}$, we determine the corresponding failure coefficients by solving \eqref{eq_hybrid_ml_map_mip} 
    using the given position and gain estimates $\pp^{(i-1)}$ and $\alpha^{(i-1)}$, i.e.,
    \begin{align}\label{eq_zetabhat_i}
    \zetabhat^{(i)} = \arg \min_{\zetab^{(i)}} \norm{	\yybrev^{(i)} - \bA^{(i)} \zetab^{(i)}  }_2^2 ~,
    \end{align}
    where $\yybrev^{(i)} \triangleq \yy - \sum_{n=1, n \notin \Iset}^{N} [\bA(\alpha^{(i-1)}, \pp^{(i-1)})]_{:,n}$, $\bA^{(i)} \triangleq \cup_{n \in \Iset} [\bA(\alpha^{(i-1)}, \pp^{(i-1)})]_{:,n} \in \complexset{T}{\lvert\Iset\rvert}$ and $\zetab^{(i)} \in \complexset{\lvert\Iset\rvert}{1}$. In \eqref{eq_zetabhat_i}, we discard the prior-related term in \eqref{eq_hybrid_ml_map_mip_obj}, approximating for high-SNR conditions, to obtain a tractable problem, as done in \eqref{eq_zetak2} and \eqref{eq_zetak2_i}. Similar to the first iteration, using the resulting $\zetabhat^{(i)}$, we update the position and channel gain estimates, re-compute \eqref{eq_zetabhat_i} and carry out these alternating steps until convergence.    
\end{itemize}

After the algorithm terminates, we will refine the mask estimate under the unit-disk constraint. Let $\widehat{\mathcal{I}}$ denote the set of failing pixels, and $\pphat$ and $\alphahat$ the position and the channel gain estimates at the end of the algorithm. Then, we formulate the following problem:
\begin{subequations} \label{cvx_for_given_idx}
\begin{align} 
	\mathop{\mathrm{min}}\limits_{\zetab\in \complexset{\lvert\widehat{\mathcal{I}}\rvert}{1}} &~~\norm{\yybrev - \widehat{\bA} \zetab  }_2^2
	 \\ ~~ \mathrm{s.t.} &~~ \left\lvert [\zetab]_n\right\rvert \leq 1, ~ \forall n \in \widehat{\mathcal{I}},
\end{align}
\end{subequations}
where  we define $\yybrev \triangleq \yy - \sum_{n=1, n \notin \widehat{\mathcal{I}}}^{N} [\bA(\alphahat, \pphat)]_{:,n}$, $\widehat{\bA} \triangleq \cup_{n \in \widehat{\mathcal{I}}} [\bA(\alphahat, \pphat)]_{:,n} \in \complexset{T}{\lvert\widehat{\mathcal{I}}\rvert}$. The problem \eqref{cvx_for_given_idx} is convex and thus can be solved using standard tools \cite{cvx}. Then, by updating the mask vector from \eqref{cvx_for_given_idx}, we can refine the final position and channel coefficient estimates by implementing \cite[Alg.~1]{Cuneyd_WCL_RIS_2022}.

The entire algorithm, called Successive-JLFD, is summarized in Algorithm~\ref{alg_successive_complete}.

\begin{algorithm}[t]
	\caption{Successive Joint Localization and Failure Diagnosis (Successive-JLFD) Algorithm to Solve \eqref{eq_hybrid_ml_map_mip}}
	\label{alg_successive_complete}
	\begin{algorithmic}[1]
		\STATE \textbf{Input:} Observation $\yy$ in \eqref{eq_y_vec_fail}, RIS phase profiles $\Phib$, convergence threshold $\varepsilon$, maximum number of alternating steps $C$ and maximum number of successive iterations $I$.
		\STATE \textbf{Output:} UE location $\pphat$, failure mask $\widehat{\mm}$, set of failing locations $\widehat{\mathcal{I}}$ and channel gain $\alphahat$.
  \STATE \textbf{Initialization:} Set $i=0$, $\mathcal{I}^{(0)} = \emptyset$ and $\mm^{(0)} = \boldone$. Compute the corresponding $\pp^{(0)}$ and $\alpha^{(0)}$ via \cite[Alg.~1]{Cuneyd_WCL_RIS_2022}.
    \STATE \textbf{Iterations:}
		\WHILE  {$i<I$}
		\STATE  Set $\ell = 0$, $i = i + 1$. $\pp^{(i,0)} = \pp^{(i-1)}$, $\alpha^{(i,0)} = \alpha^{(i-1)}$, and $\mm^{(i,0)} = \mm^{(i-1)}$.
		\WHILE {$\ell < C$}
		\STATE Solve \eqref{eq_zetak2_i} by plugging $\pp^{(i,\ell)}$, $\alpha^{(i,\ell)}$ and  $\mm^{(i,\ell)}$ from \eqref{eq_mtildek_i2}.
        \STATE Solve \eqref{eq_khat2} to determine $\khat$.
		\STATE Set $\ell = \ell + 1$.
		\IF {$\khat = 0$}
		\STATE Set $\pp^{(i)} = \pp^{(i,\ell)}$, $\alpha^{(i)} = \alpha^{(i,\ell)}$, $\mm^{(i)} = \mm^{(i,\ell)}$ and $\mathcal{I}^{(i)} = \mathcal{I}^{(i-1)}$.
  \STATE \textbf{break}
		\ELSE
            \STATE Solve \eqref{eq_zetabhat_i} to update the mask $\mm^{(i,\ell)}$ with the new failure coefficients $\zetabhat^{(i)}$.		
		\STATE For the updated mask, update $\pp^{(i,\ell)}$ and  $\alpha^{(i,\ell)}$ via \cite[Alg.~1]{Cuneyd_WCL_RIS_2022}.
		\IF{$\ell \geq C$ or $\norm{\pp^{(i,\ell)}-\pp^{(i,\ell-1)}}_2\leq \varepsilon$}
		\STATE Set $\pp^{(i)} = \pp^{(i,\ell)}$, $\alpha^{(i)} = \alpha^{(i,\ell)}$, $\mm^{(i)} = \mm^{(i,\ell)}$, $\mathcal{I}^{(i)} = \mathcal{I}^{(i-1)} \cup \{\khat\}$ and $\ell = C$.
		\ENDIF
		\ENDIF
		\ENDWHILE
		\IF {$i\geq I$}
		\STATE Terminate the algorithm with the current failing locations $\widehat{\mathcal{I}}$ and set $\widehat{\pp} = \pp^{(i)}$, $\widehat{\alpha}= \alpha^{(i)}$ and $\widehat{\mm} = \mm^{(i)}$.
     	\STATE \textbf{break}
		\ENDIF
		\ENDWHILE
  \STATE \textbf{Refinement:} Refine the mask estimate by solving \eqref{cvx_for_given_idx} under the unit-disk constraint and refine the UE location and channel gain via \cite[Alg.~1]{Cuneyd_WCL_RIS_2022}.
	\end{algorithmic}
	\normalsize
\end{algorithm}	

\section{Complexity Analysis}\label{sec:CompAnalysis}
In this section, we carry out computational complexity analysis of Algorithm~\ref{alg_ell_1} and Algorithm~\ref{alg_successive_complete}. It is assumed that the search intervals over distance, azimuth, and elevation for UE location estimation via \cite[Alg.~1]{Cuneyd_WCL_RIS_2022} are discretized into grids of size $K$.

\subsection{Complexity Analysis of Algorithm~\ref{alg_ell_1}}  The initial cost of calculating $\pp^{(0)}$ and $\alpha^{(0)}$ is given by $\mathcal{O}(TK^2N)$ \cite[Sec.~4-E]{Cuneyd_WCL_RIS_2022}. At the $\thn{i}$ iteration, for given $\alpha^{(i)}$ and $\pp^{(i)}$,  ISTA can be used to compute $\mm^{(i+1)}$ \cite{Beck_09_ISTA, TVT_2022_ISTA}. In this algorithm, we initially compute the \ac{LS} solution by calculating $\left(\bA(\alpha^{(i)}, \bp^{(i)})^\hermit\bA(\alpha^{(i)}, \bp^{(i)})\right)^{-1} \bA(\alpha^{(i)}, \bp^{(i)})^\hermit \yy$, whose computational cost is equal to $\mathcal{O}(N^2T+N^3)$. Since  $\bA(\alpha^{(i)}, \bp^{(i)})^\hermit\bA(\alpha^{(i)}, \bp^{(i)})$ has already been computed, the computational cost of each ISTA iteration is simply equal to $\mathcal{O}(N^2)$.  If $N_{\text{iter}}^{(i)}$ is the number of iterations required to achieve convergence of ISTA in the $\thn{i}$ iteration, then $\mathcal{O}(N^2T+N^3) + \mathcal{O}(N^2 N_{\text{iter}}^{(i)}) = \mathcal{O}\left(N^2(N+T+ N_{\text{iter}}^{(i)})\right)$ is the total computational cost of estimating $\mm^{(i+1)}$. In addition, for given $\mm^{(i+1)}$, $\mathcal{O}(TK^2N)$ specifies the computational cost of calculating $\pp^{(i+1)}$ and $\alpha^{(i+1)}$ \cite[Sec.~4-E]{Cuneyd_WCL_RIS_2022}. Therefore, the overall cost of the $\thn{i}$ iteration is $\mathcal{O}\left(N(TK^2 + N^2+TN+ N N_{\text{iter}}^{(i)})\right)$. 
 
 Based on the assumption that $N_{\text{iter}}$ is the maximum value of  $\{N_{\text{iter}}^{(i)}\}_i$, the overall cost of Algorithm~\ref{alg_ell_1}  is $\mathcal{O}\left(MN(TK^2 + N^2+TN+ N N_{\text{iter}})\right)$. If the search intervals for distance, azimuth and the elevation are sufficiently large, then the overall cost of Algorithm~\ref{alg_ell_1} is $\mathcal{O}\left(MN TK^2\right)$.

\subsection{Complexity Analysis of Algorithm~\ref{alg_successive_complete}}
As in Algorithm~\ref{alg_ell_1}, the initial cost of calculating $\pp^{(0)}$ and $\alpha^{(0)}$ is given by $\mathcal{O}(TK^2N)$ \cite[Sec.~4-E]{Cuneyd_WCL_RIS_2022}. At the beginning of the first iteration, $\mathcal{O}(TN)$ provides the computation cost of  $\bA(\alpha^{(0)}, \pp^{(0)})$. Then, for any alternating step $\ell$, the cost of computing $\yytilde_k^{(0)}$ for each hypothesis $k\in\{1, \ldots, N\}$ is simply $\mathcal{O}(TN)$. Then, the computational cost of \eqref{eq_zetakhat} is given by $\mathcal{O}(T)$ for any hypothesis $k$. Finally, by utilizing \eqref{eq_zetak2}, the computational cost of plugging $\mmtildee{k}$ into $\mathcal{L}(\cdot)$ is reduced to $\mathcal{O}(T)$. Since these computations must be performed for each hypothesis, we can conclude that the cost of updating the mask vector for any alternating step $\ell$ is simply equal to $\mathcal{O}(TN^2)$. As the cost of estimating the position and the channel coefficient is equal to $\mathcal{O}(TK^2N)$, the total cost of the first iteration is simply equal to  $\mathcal{O}(C(NTK^2 + TN^2))$.

Similar analyses reveal that the $\thn{i}$ iteration of Algorithm~\ref{alg_successive_complete} requires a total cost of $\mathcal{O}(C(NTK^2 + TN^2))$. Given that the maximum number of iterations is equal to $I$, the overall cost of Algorithm~\ref{alg_successive_complete} is $\mathcal{O}(CI(NTK^2 + TN^2))$. Under the condition that the search intervals are sufficiently large, the overall cost of Algorithm~\ref{alg_successive_complete} is given by $\mathcal{O}\left(CIN TK^2\right)$. Note that once the algorithm terminates, even though we solve a convex problem given by \eqref{cvx_for_given_idx} to refine the mask and position estimates, we only solve the problem once, and its complexity is negligible when compared to the rest of the algorithm.



\subsection{Complexity Comparison of Algorithm~\ref{alg_ell_1} and Algorithm~\ref{alg_successive_complete}}
Summarizing the results in the previous subsections, the overall complexities of Algorithm~\ref{alg_ell_1} and Algorithm~\ref{alg_successive_complete} are given by $\mathcal{O}\left(MN TK^2\right)$ and $\mathcal{O}\left(CIN TK^2\right)$, respectively. In the numerical simulations, we set $M$ and $C$ to be equal to each other. Thus, if $K$ is sufficiently large, the complexity of Algorithm~\ref{alg_successive_complete} is roughly $I$ times that of Algorithm~\ref{alg_ell_1}.

\section{Numerical Results}\label{sec:NumRes}
In this section, we present numerical results to evaluate the theoretical bounds derived in Sec.~\ref{sec:Problem1} and the performance of the proposed algorithms in Sec.~\ref{sec:JointLocFailDiag}. 

\subsection{Simulation Setup}\label{subsec:SimSet}
We consider an RIS with $N = 20\times 20$ elements, where the inter-element spacing is $\lambda/2$ and the area of each element is $\lambda^2/4$ \cite{Shaban2021}. The RIS aligns with the X-Y plane and is located at $\bp_{\text{RIS}}=[0~ 0~ 0]^{\trp} ~ \rm{m}$. In addition, the entries of the RIS phase configurations, $\phib_t$'s in \eqref{eq_gammab}, are generated uniformly and independently between $-\pi$ and $\pi$. Moreover, we set the number of transmissions as $T = 20$ and the carrier frequency as $f_c = 28\, \rm{GHz}$, leading to $\lambda = 10.71 \, \rm{mm}$. The BS is located at $\bp_{\text{BS}}=10\times {[1~ 1 ~ 1]^{\trp}}/{\norm{[1 ~ 1 ~ 1]}_2} ~ \rm{m}$, while the UE is located at $4\times {[1 ~ 1~ 1]^{\trp}}/{\norm{[1~ 1~ 1]}_2} ~ \rm{m}$. For convenience, we assume that $s_t = \sqrt{E_s} ~\forall t$. Also, the  SNR is defined as $\snr = \lvert\alpha\rvert^2 E_s/N_0$.

In Algorithm~\ref{alg_ell_1}, we set $\xi= 2\sqrt{\text{SNR}}$ and $M = 5$.
While implementing Algorithm~\ref{alg_successive_complete}, the number of maximum allowed iterations, $I$, is selected as $2N\pfail$. The reasoning behind this selection can be explained as follows. Since the number of pixel failures estimated by Algorithm~\ref{alg_successive_complete} is upper bounded by $I$, we need to choose $I$ such that $\Pr\{\text{Number of failures} > I\}\leq \epsilon$, where $\epsilon>0$ is a small number. More specifically,
\begin{align}\nonumber
	\Pr\{\# \text{of failures} > I\} = 1 - \sum_{m=0}^{I} (\pfail)^{m}(1-\pfail)^{N-m}  \binom{N}{m}
\end{align}
should be small. In accordance with the sparsity assumption, we consider $\pfail\in\ (0, 0.02]$ in our simulations. For these values of $\pfail$ and  for $I = 2N\pfail$, $\Pr\{\# \text{of failures} > I\} \leq 0.08$. Moreover,  we set $C= 5$ in Algorithm~\ref{alg_successive_complete} and $\varepsilon = 0.001$ in both algorithms. Finally, the search intervals for UE location estimation via \cite[Alg.~1]{Cuneyd_WCL_RIS_2022} are set to $[0, \, 50] ~ \rm{m}$ for distance and $[0, \, \pi/2]$ for azimuth/elevation and the number of grid points is taken as $K = 501$.

\subsection{Theoretical Performance Evaluation Under Pixel Failures}\label{sec_mcrb_numerical}

\begin{figure}[t]
        \begin{center}
        \subfigure[]{
			 \label{fig:theo_boundsvs_p_fail}
			 \includegraphics[width=0.45\textwidth]{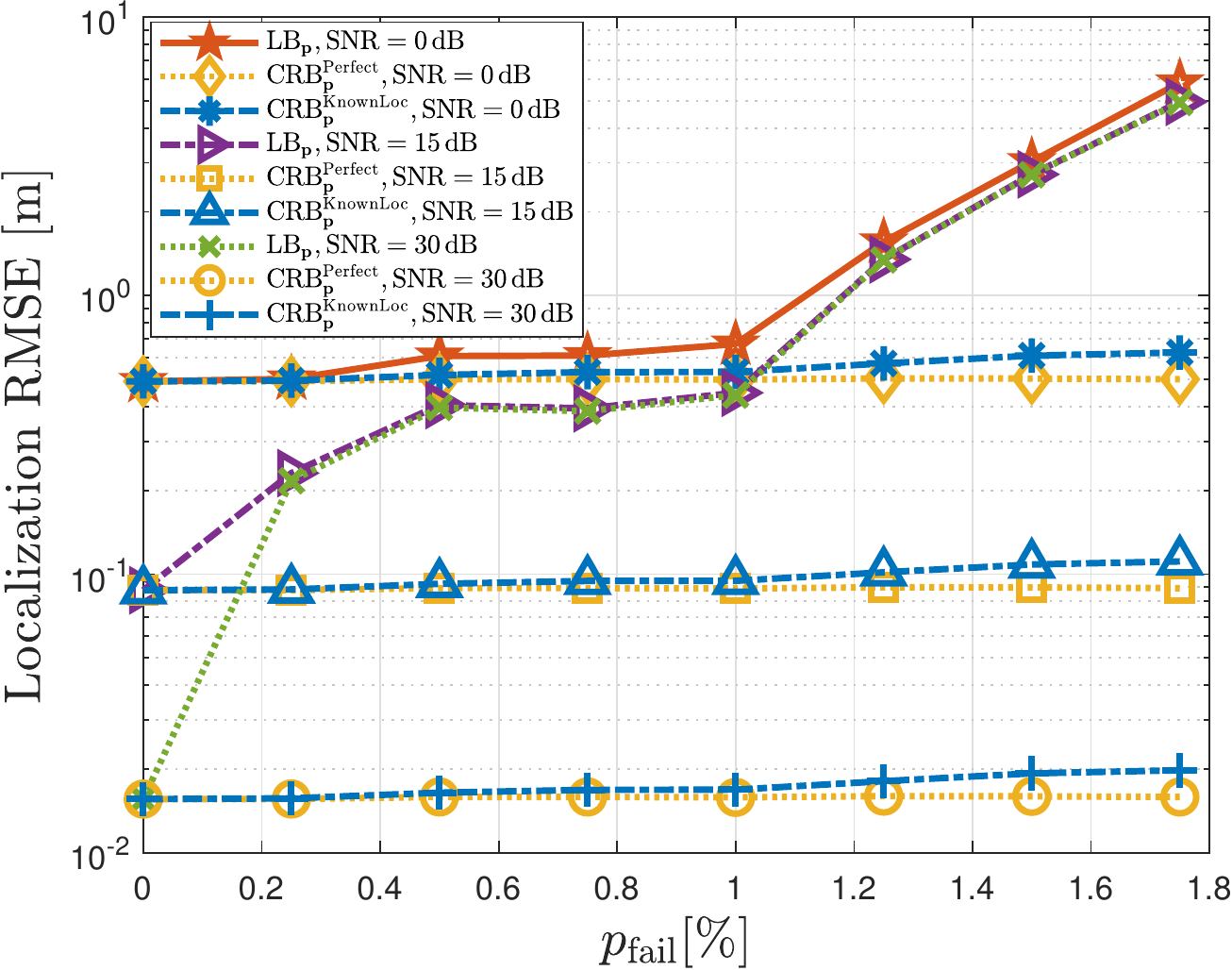}
		}
        \subfigure[]{
			 \label{fig:theo_boundsvs_SNR}
			 \includegraphics[width=0.45\textwidth]{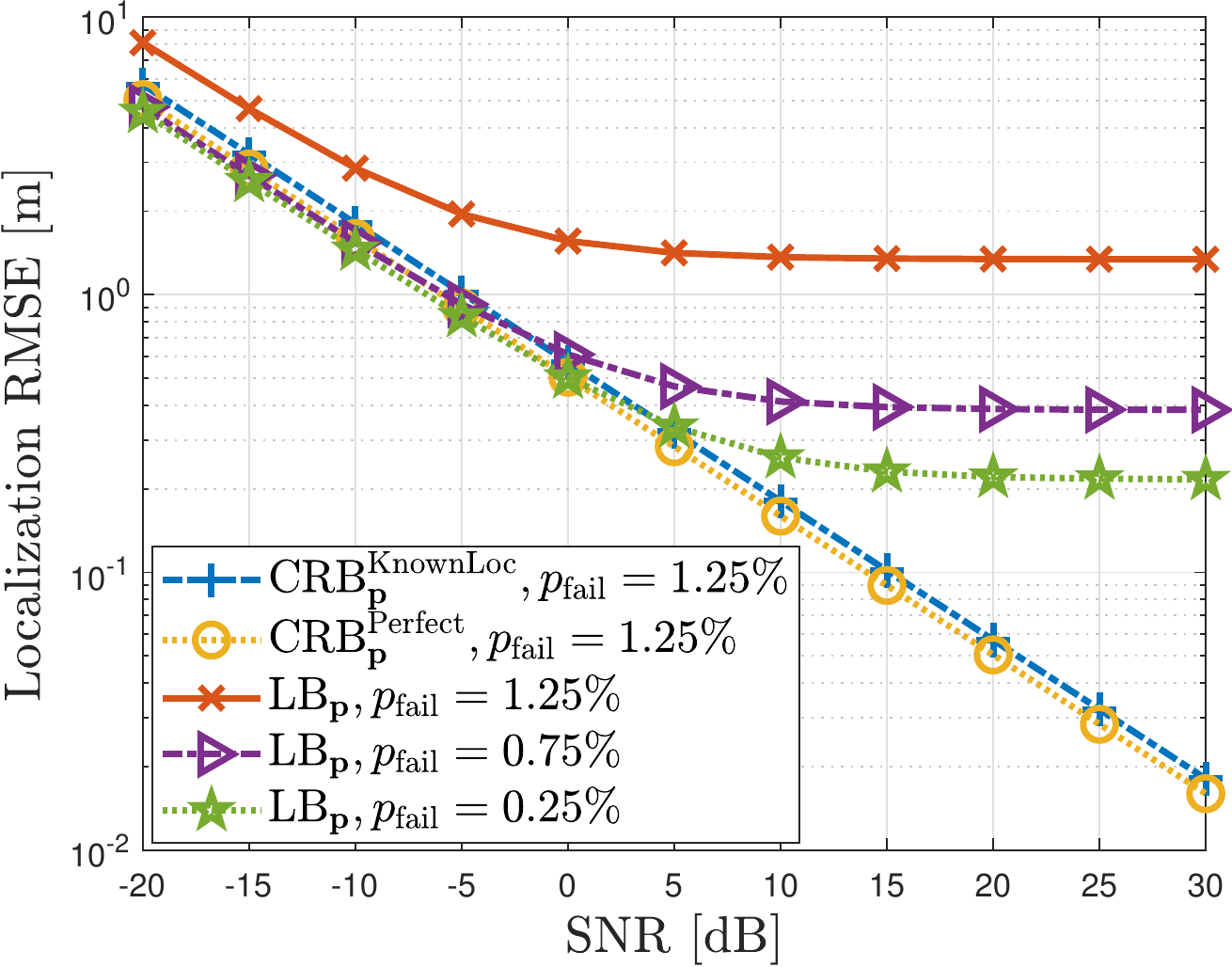}
		}
		\end{center}
		\vspace{-0.2in}
        \caption{Theoretical limits on localization \ac{RMSE} in \eqref{eq_lb_pp}, \eqref{eq_crb_perf} and \eqref{eq_crb_known} versus \subref{fig:theo_boundsvs_p_fail} $\pfail$ for various SNR values, and \subref{fig:theo_boundsvs_SNR} SNR for various $\pfail$ values. As $\crbperpp$ and $\crbknownpp$ change only marginally with respect to $\pfail$, they are plotted for a single value of $\pfail$ in \subref{fig:theo_boundsvs_SNR} to declutter the figure.} 
        \label{fig:theobounds}
        \vspace{-0.1in}
\end{figure}

In Fig.~\ref{fig:theobounds}, we report the theoretical limits on UE location estimation, derived in \eqref{eq_lb_pp}, \eqref{eq_crb_perf} and \eqref{eq_crb_known}, as a function of $\pfail$ and $\snr$. While obtaining the figures, the locations of faulty RIS elements for any given $\pfail$ are fixed at random. For instance, if $\pfail = 2\%$, we randomly assign $N\pfail = 8$ failure locations and fix them. Then, we obtain the corresponding curves by averaging over $100$ distinct failure profiles (i.e., by considering $100$ distinct realizations of $(\kappa_n,\psi_n)$ pairs in \eqref{eq_mask_m} for the fixed failure locations). 

We observe that $\crbperpp$ and $\crbknownpp$ exhibit very similar values in the considered SNR and $\pfail$ regimes. This suggests that once the locations of the failing RIS elements are known, knowing the respective failure coefficients $\zeta_n = \kappa_n e^{j \psi_n}$ brings only a marginal improvement in localization performance. Hence, the main bottleneck in RIS diagnosis with a \ac{UE} having an a-priori unknown location lies in \textit{detection} of the status (failing/functioning) of $N$ RIS elements from $T \ll N$ scalar observations, rather than \textit{estimation} of their failure coefficients. In addition, we see that the gap between the LB and CRB curves becomes larger with increasing $\pfail$ (i.e., as the mismatch between the true model in \eqref{eq_y_vec_fail_true} and the assumed model in \eqref{eq_y_vec_fail_assumed} increases), as expected. When the UE is unaware of pixel failures, severe performance degradations can occur, especially at high SNRs (e.g., more than two orders-of-magnitude accuracy loss at $\snr = 30 \, \rm{dB}$ for $\pfail > 1\%$). Therefore, even with small percentage of failures, employing failure-agnostic algorithms might lead to significant penalties in localization. Moreover, looking at the LB curves in Fig.~\ref{fig:theo_boundsvs_SNR}, we observe that the localization performance saturates above a certain SNR level, reaching an SNR-independent bias value quantified by the second term in \eqref{eq_lb}. This results from the mismatch between the true and assumed models, i.e., the price paid due to failures being ignored. Overall, we can conclude that RIS pixel failures can significantly degrade the localization performance, which necessitates the design of effective algorithms to mitigate their impact.

The high sensitivity of localization to pixel failures can be attributed to the fact that in the considered narrowband \ac{SISO} setup with \ac{LoS} blockage, information on \ac{UE} location $\pp$ derives only from the phase shifts across the \ac{RIS} elements, as seen from the \ac{NF} steering vector $\aab(\pp)$ in \eqref{eq_nf_steer} and the associated observations in \eqref{eq_yt}. As pixel failures distort the phase profiles as specified in \eqref{eq_gammab}, the information that can be extracted from a set of scalar observations at the \ac{UE} becomes quite inaccurate.

\subsection{Performance of Algorithm~\ref{alg_ell_1} and Algorithm~\ref{alg_successive_complete}}\label{sec_perf_alg}

\begin{figure}[t]
        \begin{center}
        \subfigure[]{
			 \label{fig_rmse_pfail_2}
			 \includegraphics[width=0.45\textwidth]{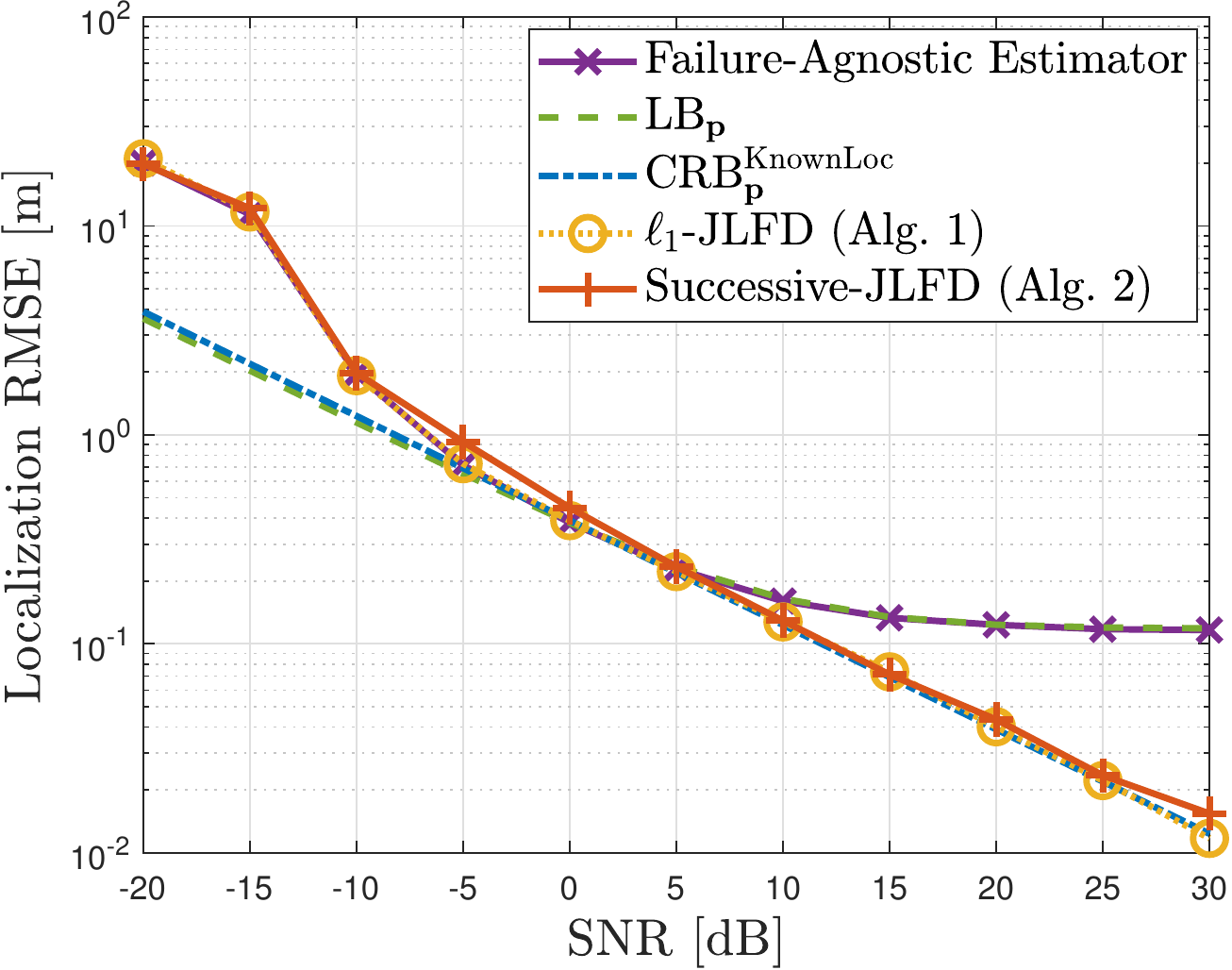}
		}
        \subfigure[]{
			 \label{fig_rmse_pfail_4}
			 \includegraphics[width=0.45\textwidth]{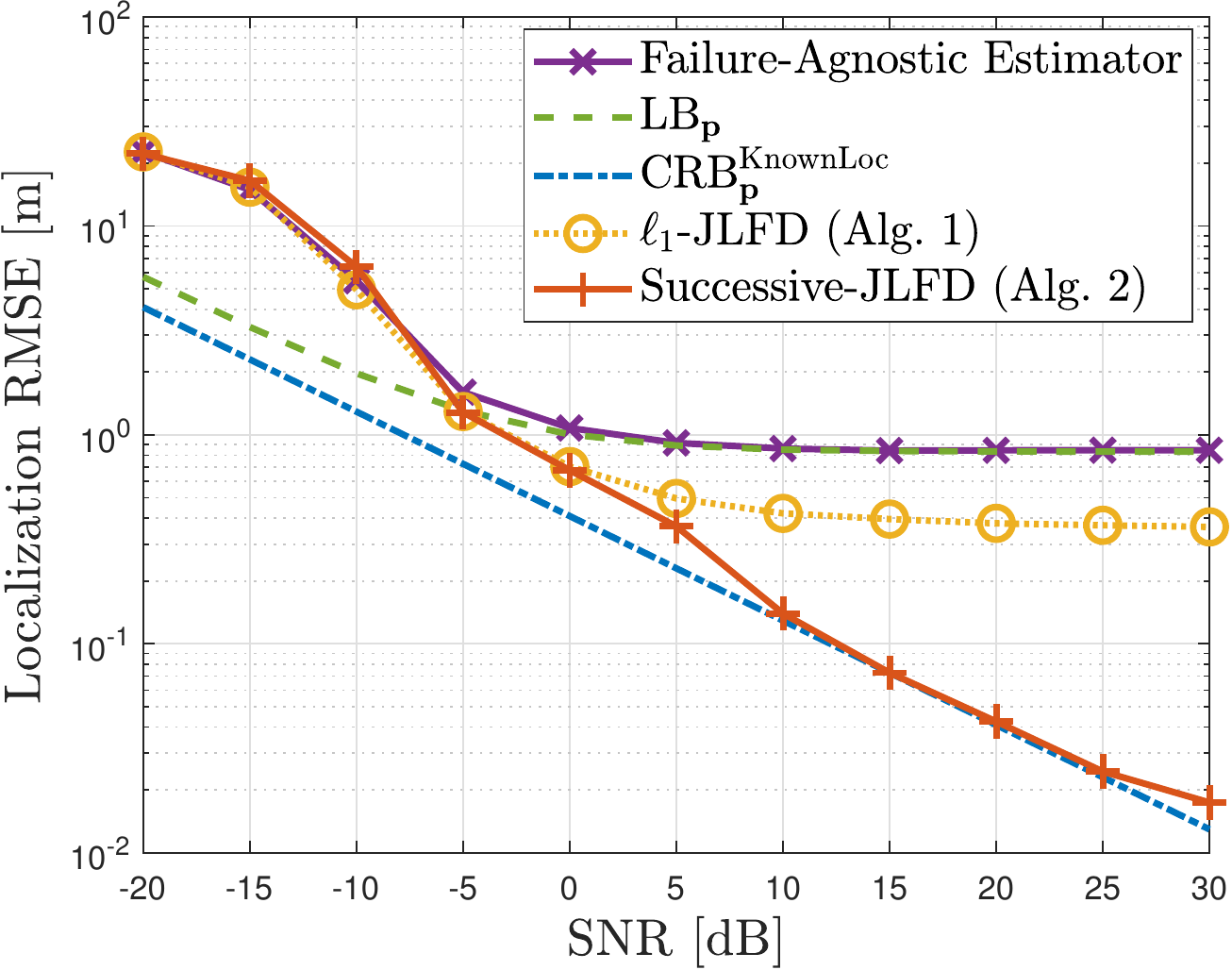}
		}
		\end{center}
		\vspace{-0.2in}
        \caption{\rev{Localization \acp{RMSE} obtained by the failure-agnostic MS-unbiased estimator $\etabhat(\yy)$ in \eqref{eq_lb_ms_unb} and the proposed algorithms, along with the theoretical bounds in \eqref{eq_lb_pp} and \eqref{eq_crb_known}, with respect to SNR for \subref{fig_rmse_pfail_2} $\pfail = 0.5 \%$, and \subref{fig_rmse_pfail_4} $\pfail = 1 \%$.}}
        \label{fig_rmse_pfail_algs}
        \vspace{-0.1in}
\end{figure}

 \begin{figure}[t]
	\centering
	\includegraphics[width = 0.9 \linewidth]{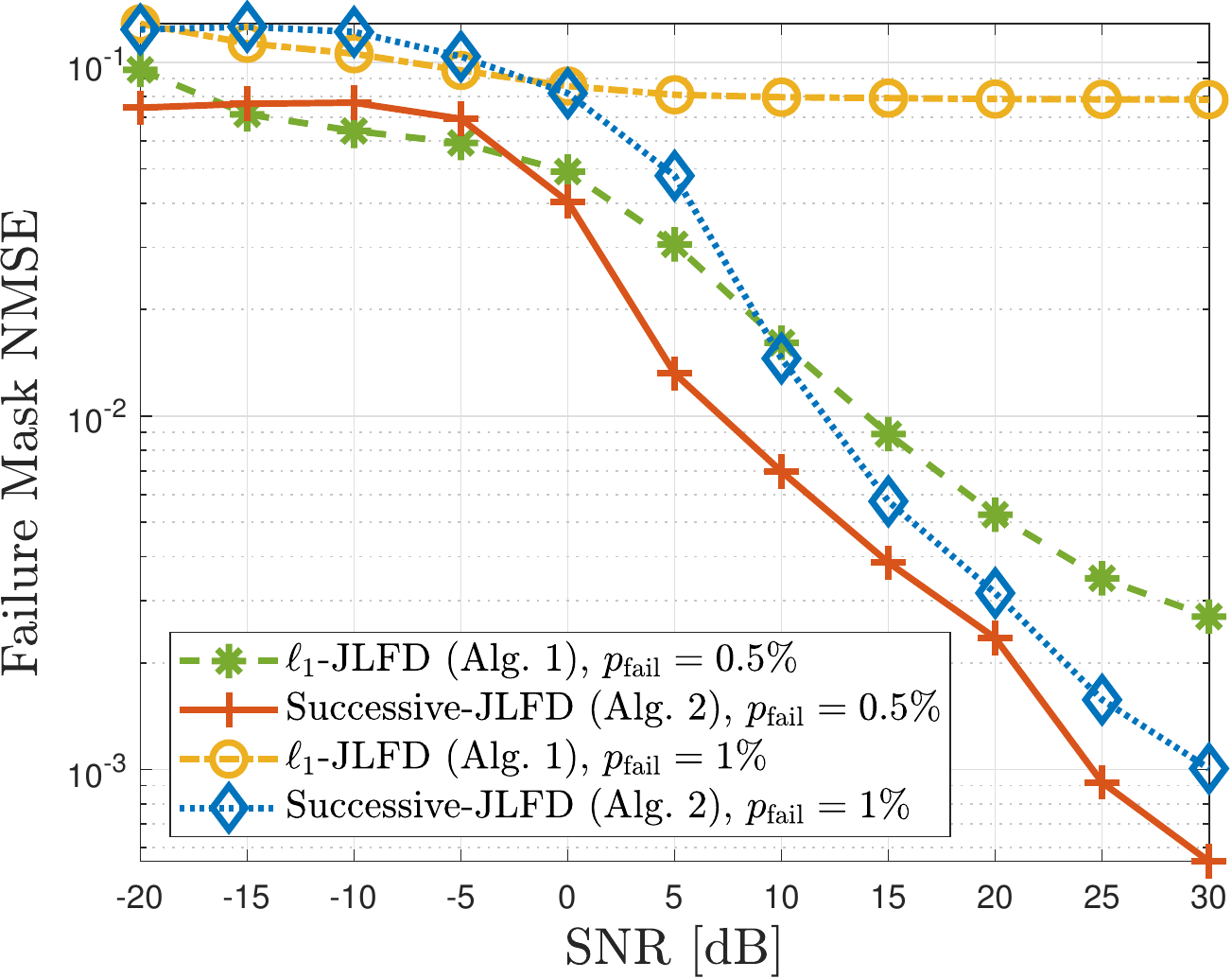}
	\caption{\rev{Mask recovery performances (mask NMSE in \eqref{eq_nmse}) of Algorithm~\ref{alg_ell_1} and Algorithm~\ref{alg_successive_complete} versus SNR for $\pfail = 0.5 \%$ and $\pfail = 1 \%$.}}
	\label{fig:NMSEvsSNR_NF_2_4}
 \vspace{-0.1in}
\end{figure}

 In this part, we examine the \rev{localization \ac{RMSE} performance of the following algorithms:}
 \rev{\begin{itemize}
     \item \textbf{Failure-Agnostic Estimator:} The failure-agnostic estimator $\etabhat(\yy)$ in \eqref{eq_lb_ms_unb}, corresponding to the case where the UE is unaware of pixel failures. This will be used as a benchmark to demonstrate performance gains of the proposed JLFD algorithms.
     \item \textbf{$\ell_1$-JLFD:} The proposed JLFD algorithm based on $\ell_1$ regularization in Algorithm~\ref{alg_ell_1}.
     \item \textbf{Successive-JLFD:} The proposed successive failure detection and UE localization algorithm in Algorithm~\ref{alg_successive_complete}.
 \end{itemize}}
 \rev{We will also evaluate the} mask recovery performances of Algorithm~\ref{alg_ell_1} and Algorithm~\ref{alg_successive_complete}. 
 In Monte Carlo simulations, for a given $\pfail$ value, $\lfloor N \pfail\rfloor$ pixels out of $N$ are randomly chosen as failing and the corresponding failure coefficients represented by $(\kappa_n, \psi_n)$ are randomly generated. Given this single realization of the failure mask $\mm$, the algorithm performances are evaluated by averaging over \rev{$200$} distinct noise realizations. In addition, the mask recovery performance is characterized through the \ac{NMSE}, defined as 
 \begin{equation} \label{eq_nmse}
	\nmse = \frac{\norm{\mmhat-\mm}^2_2}{\norm{\mm}^2_2} ~,
\end{equation}
where $\mmhat$ denotes the mask estimate.
 
 

\subsubsection{\rev{Performance with respect to SNR}}
 In Fig.~\ref{fig_rmse_pfail_algs}, we show the localization \ac{RMSE}s achieved by the MS-unbiased estimator in \eqref{eq_lb_ms_unb} (i.e., the \textit{failure-agnostic benchmark}), Algorithm~\ref{alg_ell_1} and Algorithm~\ref{alg_successive_complete} as a function of SNR, along with the theoretical bounds, for $\pfail = 0.5 \%$ and $\pfail = 1 \%$. The corresponding mask \ac{NMSE} performances of Algorithm~\ref{alg_ell_1} and Algorithm~\ref{alg_successive_complete} are illustrated in  Fig.~\ref{fig:NMSEvsSNR_NF_2_4}. First, we see that the failure-agnostic MS-unbiased estimator in \eqref{eq_lb_ms_unb} achieves the corresponding LB asymptotically at high SNRs, which corroborates the MCRB analysis in Sec.~\ref{sec_mcrb}. In addition, it is observed that Algorithm~\ref{alg_successive_complete} significantly outperforms the failure-agnostic estimator and attains the corresponding CRB at an SNR of $0$ and $10 ~ \rm{dB}$ for $\pfail = 0.5 \%$ and $\pfail = 1 \%$, respectively, which demonstrates the effectiveness of the proposed successive \ac{JLFD} strategy in recovering failure-induced performance degradations. Hence, Algorithm~\ref{alg_successive_complete} can successfully yield accurate estimates of \ac{UE} location in the presence of pixel failures and provide performance achievable under perfect knowledge of the failure mask $\mm$ in \eqref{eq_gammab}, corresponding to a perfectly calibrated \ac{RIS}. Moreover, by comparing Fig.~\ref{fig_rmse_pfail_2} and Fig.~\ref{fig_rmse_pfail_4}, it appears that Algorithm~\ref{alg_ell_1} exhibits performance similar to that of Algorithm~\ref{alg_successive_complete} for $\pfail = 0.5 \%$, while it fails to achieve the theoretical limit for $\pfail = 1 \%$ and reaches a plateau in localization accuracy above a certain SNR level. This results from limited usage of the failure statistics in the $\ell_1$-regularization approach, in contrast with full exploitation of the statistics in the successive-\ac{JLFD} approach. It is worth noting that the advantage of the $\ell_1$-regularization strategy lies in its low computational complexity, as investigated in Sec.~\ref{sec:CompAnalysis}\footnote{Run-time analysis during the simulations show that Algorithm~\ref{alg_ell_1} and Algorithm~\ref{alg_successive_complete} take $3.7$ and $12.9$ times longer than the failure-agnostic estimator, respectively.}. Finally, the mask \ac{NMSE} performances in Fig.~\ref{fig:NMSEvsSNR_NF_2_4} confirm the localization \ac{RMSE} results in Fig.~\ref{fig_rmse_pfail_algs}. Namely, Algorithm~\ref{alg_ell_1} cannot satisfactorily recover the failure mask for $\pfail = 1 \%$, leading to gaps between the \ac{RMSE} and the \ac{CRB} in Fig.~\ref{fig_rmse_pfail_4}, while the mask \ac{NMSE} of Algorithm~\ref{alg_successive_complete} decreases consistently with SNR for both $\pfail$ values.

 \subsubsection{\rev{Performance with respect to UE Distance}}\label{sec_perf_dist}
\begin{figure}
	\centering
	\includegraphics[width=0.9\linewidth]{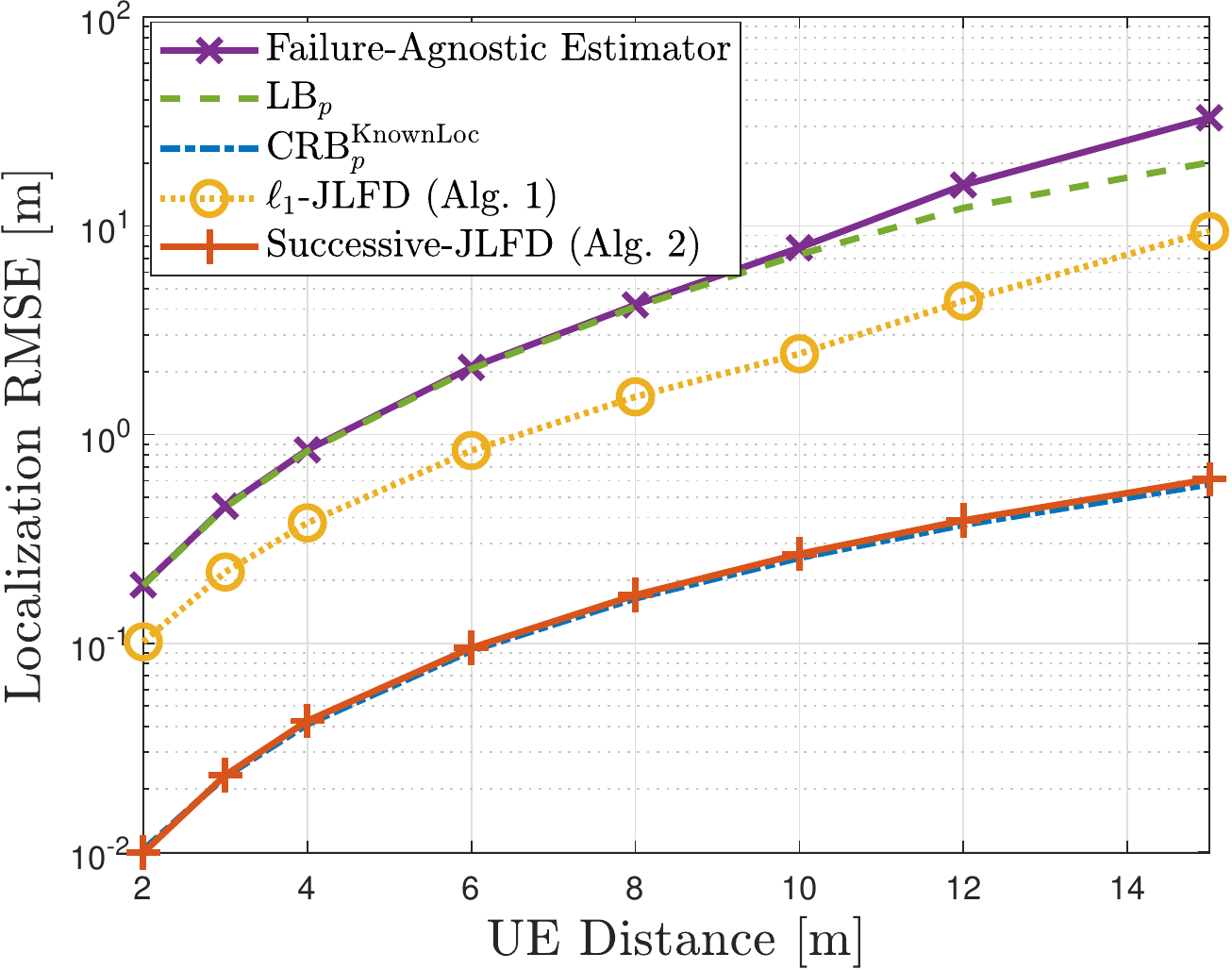}
	\caption{\rev{Localization \acp{RMSE} obtained by the failure-agnostic MS-unbiased estimator $\etabhat(\yy)$ in \eqref{eq_lb_ms_unb} and the proposed algorithms, along with the theoretical bounds in \eqref{eq_lb_pp} and \eqref{eq_crb_known}, with respect to UE distance, where $\pfail = 1 \%$ and $\snr = 20 \, \rm{dB}$.}}
	\label{fig_rmse_distance_paper}
\end{figure}
\rev{In this part, we evaluate the performance as a function of \ac{UE} distance from the \ac{RIS} to demonstrate the impact of wavefront curvature on accuracy. Fig.~\ref{fig_rmse_distance_paper} shows the localization \ac{RMSE} obtained by the considered schemes with respect to UE distance $d$ for $\pfail = 1 \%$ and $\snr = 20 \, \rm{dB}$, where the UE location is given by $d\times {[1 ~ 1~ 1]^{\trp}}/{\norm{[1~ 1~ 1]}_2} ~ \rm{m}$. It is observed that as the UE moves away from the RIS, the wavefront curvature becomes less pronounced, leading to a degradation in localization performance. This is expected, as the only information available for localization in \eqref{eq_yt} is the wavefront curvature, which is manifested in the \ac{NF} steering vector \eqref{eq_nf_steer}. Moreover, Fig.~\ref{fig_rmse_distance_paper} corroborates the validity of the proposed algorithms both in the \ac{NF} and \ac{FF} region of the \ac{RIS}, i.e., both within and beyond the Fraunhofer distance $d_{\rm{F}} = 2 D^2/\lambda = 3.86~{\rm{m}}$.}



 \begin{figure}[t]
	\centering
	\includegraphics[width = 0.9 \linewidth]{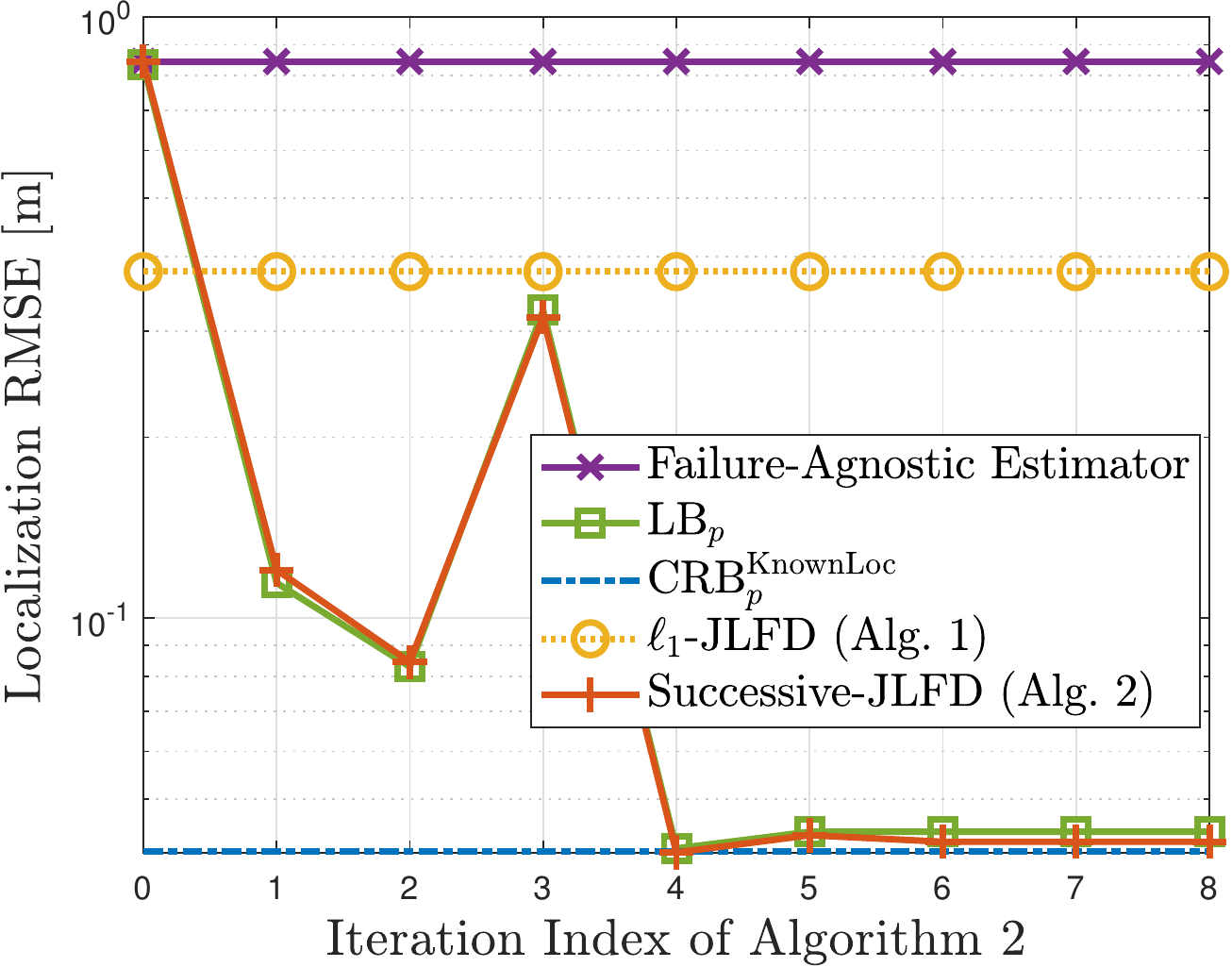}
	\caption{\rev{Evolution of localization \acp{RMSE} obtained by Algorithm~\ref{alg_successive_complete} over successive iterations, along with the theoretical bounds in \eqref{eq_lb_pp} and \eqref{eq_crb_known}, the \acp{RMSE} of Algorithm~\ref{alg_ell_1} and the failure-agnostic estimator $\etabhat(\yy)$ in \eqref{eq_lb_ms_unb}, for SNR = $20 \, \rm{dB}$ and $\pfail = 1 \%$. At each iteration, which corresponds to detection of a single failure by Algorithm~\ref{alg_successive_complete}, $\lbpp$ in \eqref{eq_lb_pp} is computed by selecting the assumed model in \eqref{eq_p_assumed} such that the UE assumes the presence of pixel failures detected up to and including the current iteration.}}
	\label{fig:RMSEvsIter_NF_4}
\end{figure}

 \begin{figure}[t]
	\centering
	\includegraphics[width = 0.9 \linewidth]{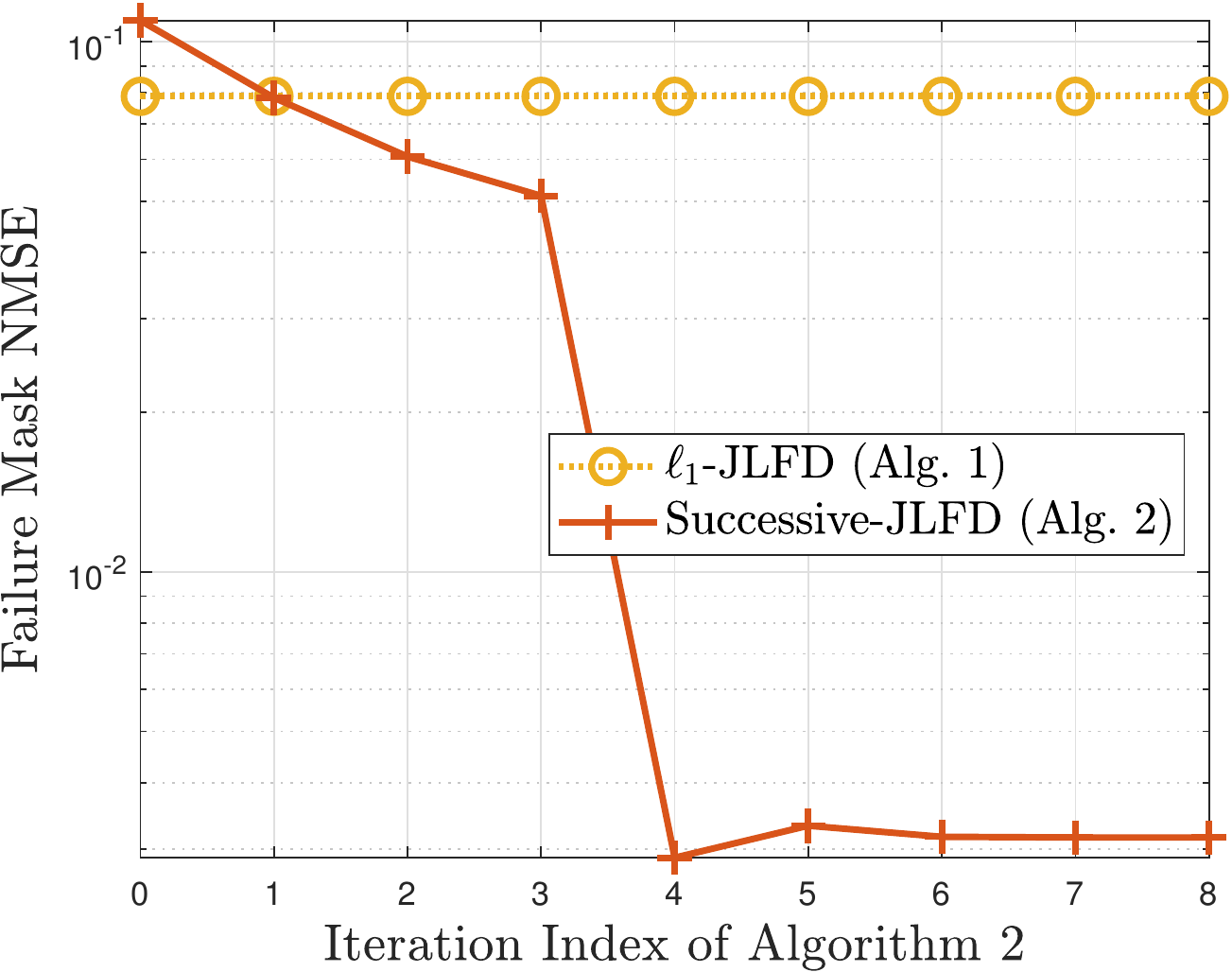}
	\caption{\rev{Evolution of mask recovery performances (mask NMSE in \eqref{eq_nmse}) of Algorithm~\ref{alg_successive_complete} over successive iterations, along with the mask NMSE of Algorithm~\ref{alg_ell_1}, for SNR = $20 \, \rm{dB}$ and $\pfail = 1 \%$.}}
	\label{fig:NMSEvsIter_NF_4}
\end{figure}

 \begin{figure*}[t]
	\centering
	\includegraphics[width = \linewidth]{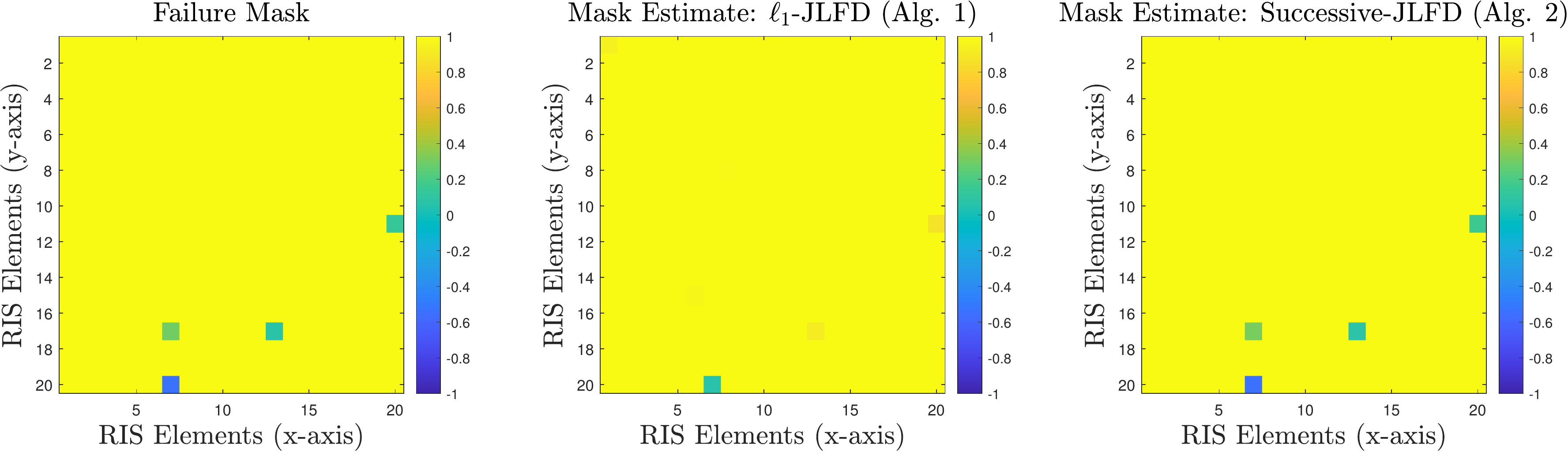}
	\caption{Illustration of failure mask $\mm$ in \eqref{eq_mask_m} over the 2-D \ac{RIS} plane, together with the mask estimates $\mmhat$ obtained by Algorithm~\ref{alg_ell_1} and Algorithm~\ref{alg_successive_complete}, for SNR = $20 \, \rm{dB}$ and $\pfail = 1 \%$ (the real parts of the masks are presented).}
	\label{fig:mask_diagnosis}
\end{figure*}


\subsubsection{\rev{Convergence Behavior}}\label{sec_conv_beh}
To investigate the convergence behavior of Algorithm~\ref{alg_successive_complete}, we plot in Fig.~\ref{fig:RMSEvsIter_NF_4} and Fig.~\ref{fig:NMSEvsIter_NF_4} the evolution of localization and mask recovery performances of Algorithm~\ref{alg_successive_complete} over successive iterations for SNR = $20 \, \rm{dB}$ and $\pfail = 1 \%$, together with $\crbknownpp$, $\lbpp$ and the localization \acp{RMSE} of the failure-agnostic benchmark and Algorithm~\ref{alg_ell_1}. It can be seen from Fig.~\ref{fig:RMSEvsIter_NF_4} that starting from the location estimate of the failure-agnostic estimator, the proposed successive-\ac{JLFD} algorithm can converge to the CRB by detecting pixel failures one-by-one at each iteration, leading to a successful progressive calibration procedure. We also emphasize that the RMSE of the successive-\ac{JLFD} algorithm attains the corresponding LB at each iteration, which indicates the \rev{superiority} of the proposed successive detection approach. In addition, it should not be surprising that the LB converges to the CRB as more failures are detected due to decreasing mismatch between the true and assumed models in \eqref{eq_p_true} and \eqref{eq_p_assumed}, respectively. \rev{However, non-monotonicity in localization performance as a function of the number of failures can still arise, as observed from Fig.~\ref{fig:RMSEvsIter_NF_4}. This behavior can be explained as follows: Localization in the considered \ac{NLoS} \ac{SISO} scenario relies purely on location-dependent RIS phase shifts (see \eqref{eq_yt} and \eqref{eq_nf_steer}), resulting in high sensitivity to pixel failures and their spatial configuration. This implies that in some rare cases, increasing the number of failures might lead to better performance since certain RIS phase changes can indeed be conducive to localization. In this respect, the non-monotonic behavior of the LB in Fig.~\ref{fig:RMSEvsIter_NF_4} across the iterations (i.e., as the number of detected/calibrated failures increases) stems from the fact that failures may counteract or reinforce each other depending on the specific failure mask realization, leading to non-monotonic localization performance with respect to the number of failures.} 

\rev{Regarding Fig.~\ref{fig:RMSEvsIter_NF_4}}, we \rev{also} note that after the $\thn{4}$ iteration (corresponding to detection of $N \pfail = 4$ pixels), Algorithm~\ref{alg_successive_complete} terminates and does not declare any new pixel as failing, and its performance coincides with the theoretical bound. This can also be observed from Fig.~\ref{fig:NMSEvsIter_NF_4}, where the mask \ac{NMSE} of Algorithm~\ref{alg_successive_complete} reaches its minimum at the $\thn{4}$ iteration. Overall, the results reveal the capability of Algorithm~\ref{alg_successive_complete} to carry out UE localization and RIS diagnosis simultaneously, for the challenging \ac{SISO} scenario under consideration. 




\subsubsection{\rev{Illustrative Example for Failure Diagnosis}}
To provide an illustrative example of failure mask recovery, Fig.~\ref{fig:mask_diagnosis} depicts the true failure mask and the instances of the estimated ones from Algorithm~\ref{alg_ell_1} and Algorithm~\ref{alg_successive_complete} for SNR = $20 \, \rm{dB}$ and $\pfail = 1 \%$. We observe that Algorithm~\ref{alg_successive_complete} performs significantly better than Algorithm~\ref{alg_ell_1} in detecting the pixel failures, and its coefficient estimates are very close to the true failure mask, in compliance with the aforementioned results and discussions.  

\subsection{\rev{Performance in the Presence of Rician Fading}}\label{sec_perf_rician}
\rev{In this part, we investigate the localization performance under Rician fading in the RIS-UE channel of our RIS-aided \ac{DL} localization scenario in Fig.~\ref{fig:setup} \cite{rayleighRician_CommL_2023,Rician_PhaseOpt_TCOM_2022}. To this end, we adopt the commonly used model where the BS-RIS link is \ac{LoS} \cite{RIS_opt_Rician_2023,IRS_Statistical_2021} and the RIS-UE link is modeled as Rician \cite{peng2021ris,Rician_PhaseOpt_TCOM_2022}. In this case, the signal model in \eqref{eq_yt} can be adapted as \cite{RIS_opt_Rician_2023,IRS_Statistical_2021,peng2021ris,Rician_PhaseOpt_TCOM_2022}
\begin{equation}\label{eq_yt_rician}
	y_t = \hhru^\trp \diag(\gammab_t) \hhbr s_t + n_t\,,
\end{equation}
where $\hhbr \in \complexset{N}{1}$ and $\hhru \in \complexset{N}{1}$ denote the BS-RIS \ac{LoS} channel and the RIS-UE Rician fading channel, respectively, which can be written as
\begin{align}\label{eq_hhbr}
    \hhbr &= \alphabr  \aab(\ppbs) \in \complexset{N}{1} \,, \\ \label{eq_hhru}
    \hhru &= \alpharu \left( \sqrt{\frac{K}{K+1}} \aab(\pp) +  \sqrt{\frac{1}{K+1}} \hhrut \right) \in \complexset{N}{1} \,.
\end{align}
Here, $\alphabr \in \complexsett$ and $\alpharu \in \complexsett$ represent the large-scale fading amplitude coefficients of the the BS-RIS and the RIS-UE link, respectively, $\hhrut \in \complexset{N}{1}$ represents the \ac{NLoS} component of the RIS-UE channel with $\hhrut \sim \mtCN(\boldzero, \Imatrix)$, and $K$ is the Rician factor \cite{Rician_PhaseOpt_TCOM_2022}.}
\revv{As seen from \eqref{eq_hhru}, Rician fading covers Rayleigh fading as a special case when $K \to 0$. In this case, the \ac{LoS} component $\aab(\pp)$ in \eqref{eq_hhru} vanishes and the information on the UE location $\pp$ is lost, meaning that the UE cannot be localized\footnote{\label{fn_rayleigh}\revv{In the presence of Rayleigh fading in the RIS-UE channel, range estimation can be performed under additional assumptions regarding the path-loss model for $\alpharu$ \cite{rayleighLoc_2021}. However, localization is still not possible in this scenario due to the absence of angular information.}}.}

\begin{figure}
	\centering
	\includegraphics[width=0.9\linewidth]{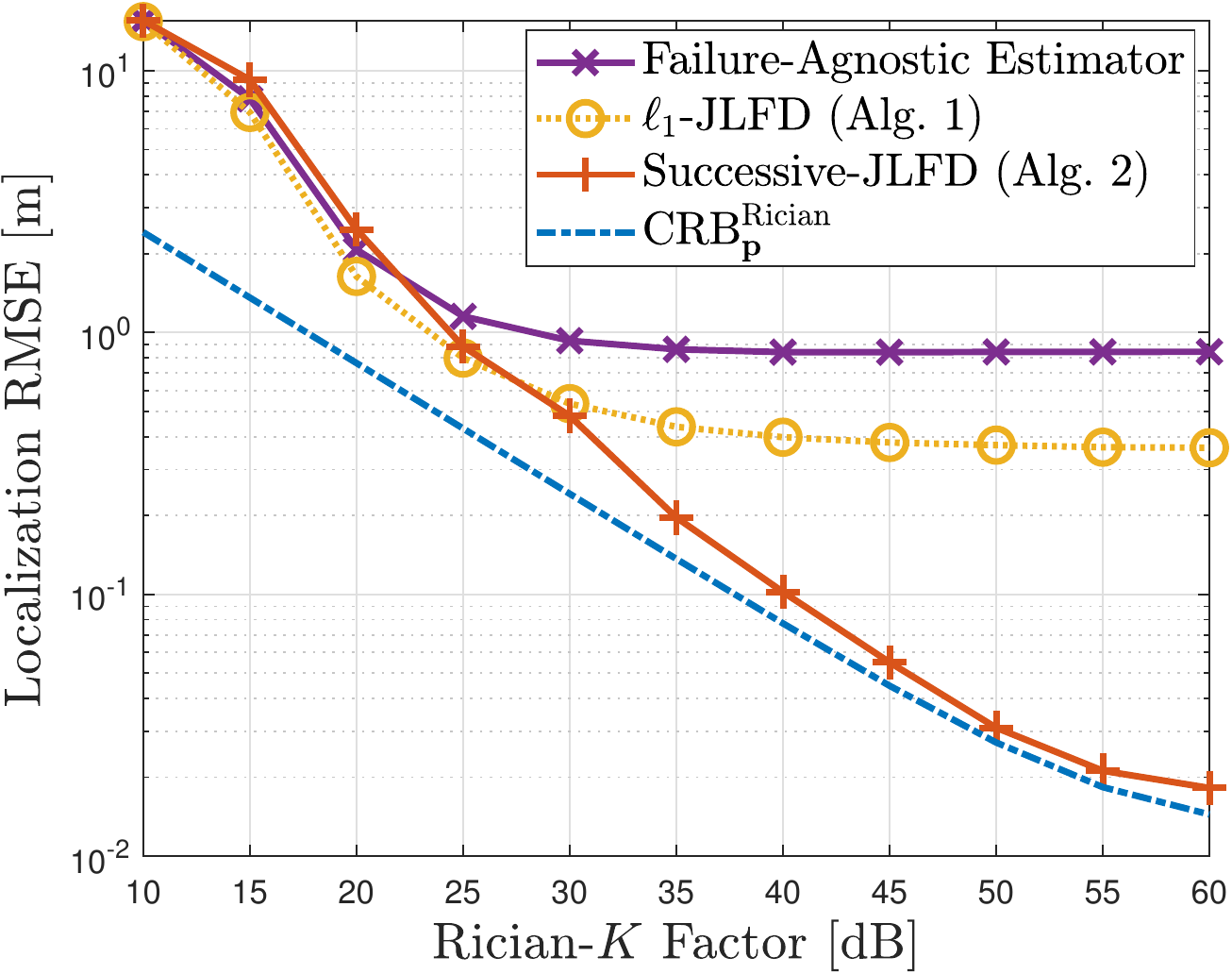}
	\caption{\rev{Localization \acp{RMSE} obtained by the failure-agnostic MS-unbiased estimator $\etabhat(\yy)$ in \eqref{eq_lb_ms_unb} and the proposed algorithms, along with the theoretical bound derived for the Rician model in \eqref{eq_yt_rician}, with respect to the Rician $K$ factor where $\pfail = 1 \%$ and $\snr = 30 \, \rm{dB}$.}}
	\label{fig_rmse_rician_paper}
\end{figure}

\rev{To evaluate how the Rician fading model in \eqref{eq_yt_rician} affects the localization performance, we plot in Fig.~\ref{fig_rmse_rician_paper} the CRB corresponding to \eqref{eq_yt_rician}, denoted by $\crbricianpp$, and the localization \ac{RMSE} results obtained by the considered schemes over $200$ Monte Carlo realizations with respect to $K$. Looking at the CRB curve in Fig.~\ref{fig_rmse_rician_paper}, we observe that the accuracy degrades with decreasing $K$ as expected since the RIS-UE channel becomes less \ac{LoS}-dominant, conveying less position information. In addition, Algorithm~\ref{alg_successive_complete} can exhibit performance very close to the CRB while the failure-agnostic estimator and Algorithm~\ref{alg_ell_1} reach a plateau in localization accuracy as $K$ increases. In the low-$K$ regime, all the three approaches achieve similar performance since the main limiting factor is $K$ (i.e., the effect of pixel failures and the ability to combat them are overshadowed by the presence of significant \ac{NLoS} components). In the high-$K$ regime, the effect of pixel failures becomes dominant and the proposed successive-JLFD algorithm (Algorithm~\ref{alg_successive_complete}) significantly outperforms the failure-agnostic benchmark and Algorithm~\ref{alg_ell_1} (similar to the results with respect to SNR in Fig.~\ref{fig_rmse_pfail_algs}).}

\section{Concluding Remarks}
In this paper, we have addressed the problem of RIS-aided localization under pixel failures and investigated the impact of such failures on localization accuracy by conducting a comparative analysis of \ac{MCRB}-based theoretical limits and standard \ac{CRB}. To counteract the effect of failures, we have proposed two algorithms, namely $\ell_1$-\ac{JLFD} and Successive-JLFD, for joint localization and mask recovery. Simulation results have offered valuable insights into the \textit{sensitivity of localization to pixel failures}. In particular, we have observed that accuracy degradation caused by pixel failures can be significant (reaching as high as two orders-of-magnitude loss at high SNRs) even in the presence of small percentage of failing elements. This can be explained by noting that localization in the considered \ac{NF} setup with \ac{LoS} blockage relies completely on phase shifts across the \ac{RIS} elements and even small number of failures can severely distort the received signal at the \ac{UE}, preventing accurate location estimation. Remarkably, the proposed algorithms can drastically reduce the localization errors compared to the failure-agnostic estimator, and asymptotically attain the corresponding theoretical limits as the SNR increases, especially for Successive-JLFD. Potential future work includes investigation of the effect of spatial distribution of failures on localization accuracy and related mitigation strategies.


\appendices

\section{PDF of Failure Coefficients}\label{app_pdf}
Let us define the random variable $\zeta = \kappa e^{j \psi}$, where $\kappa\sim\mathcal{U}(0,1)$ and $\psi\sim\mathcal{U}(-\pi,\pi)$ are independent random variables, by dropping the RIS element index $n$.  There is a one-to-one mapping between $(\kappa,\psi)$ and $(\zetar, \zetai)$, where $\zetar = \realp{\zeta} = \kappa\cos(\psi)$ and $\zetai = \imp{\zeta} =\kappa\sin(\psi)$. Hence, for given $\zetar, \zetai$, we can determine $\kappa, \psi$ as follows:
\begin{align}
    \kappa &= \sqrt{\zetar^2 + \zetai^2}, \\
\psi &= 
  \begin{cases}
    \tan^{-1} \left(\frac{\zetai}{\zetar}\right), & \text{if } \zetar \geq 0, \\
    \tan^{-1} \left(\frac{\zetai}{\zetar}\right)-\pi, & \text{if } \zetar < 0, \zetai < 0,\\
    \tan^{-1} \left(\frac{\zetai}{\zetar}\right)+ \pi, & \text{if } \zetar < 0, \zetai \geq 0.
  \end{cases}
\end{align}
Then, the Jacobian matrix can be calculated as
\begin{equation}
    \mathbf{J} = \begin{bmatrix}
    \frac{\zetar}{\sqrt{\zetar^2 + \zetai^2}}       & \frac{\zetai}{\sqrt{\zetar^2 + \zetai^2}}   \\
    \frac{-\zetai}{{\zetar^2 + \zetai^2}}      &    \frac{\zetar}{{\zetar^2 + \zetai^2}}
\end{bmatrix} ~.
\end{equation}
This implies
\begin{equation}
    f_{\zeta}(\zetar,\zetai) = f_{\kappa}(\kappa) f_{\psi}(\psi) \abss{\mathbf{J}} = \frac{1}{2\pi \sqrt{\zetar^2  + \zetai^2}}
\end{equation}
for $\zetar^2 + \zetai^2 \leq 1$. In a more compact form,
\begin{align}\label{eq_pdf_zeta}
     f_{\zeta}(\zeta) = \begin{cases}
       \frac{1}{2\pi \abss{\zeta}}, &~~  \abss{\zeta} \leq 1
       \\
       0, &~~ \textrm{otherwise}
     \end{cases} ~.
\end{align}

\vspace{-0.2in}
\section{Computation of \eqref{eq_hybrid_ml_map_mip_k_i}}\label{app_comp_2nd}

The second term in \eqref{eq_hybrid_ml_map_mip_k_i} can be obtained from \eqref{eq_pdf_mn} and \eqref{eq_mtildek_i} as follows:
\begin{align} \nonumber
    &\sum_{n=1}^{N} \log f_{m_n}([\mmtildee{k}]_n) =
    (N-1-\lvert\mathcal{I}^{(i-1)}\rvert) \log (1-\pfail)
    \\ \nonumber
    & ~~+ (\lvert\mathcal{I}^{(i-1)}\rvert + 1) \log \pfail  + \sum_{n \in \mathcal{I}^{(i-1)}} \log f_{\zeta_n}([\mm^{(i-1)}]_{n}) \\ \label{eq_logfm_i}
    & ~~+  \log f_{\zeta_k}(\zeta_k)
\end{align}
for $k \geq 1$, and
\begin{align} \label{eq_logfm_i2}
    &\sum_{n=1}^{N} \log f_{m_n}([\mmtildee{k}]_n) =
    (N-\lvert\mathcal{I}^{(i-1)}\rvert) \log (1-\pfail)
    \\ \nonumber
    & + \lvert\mathcal{I}^{(i-1)}\rvert \log \pfail  + \sum_{n \in \mathcal{I}^{(i-1)}} \log f_{\zeta_n}([\mm^{(i-1)}]_{n})  
\end{align}
for $k=0$.

\bibliographystyle{IEEEtran}
\bibliography{bibfile}

\begin{thebibliography}{10}
\providecommand{\url}[1]{#1}
\csname url@samestyle\endcsname
\providecommand{\newblock}{\relax}
\providecommand{\bibinfo}[2]{#2}
\providecommand{\BIBentrySTDinterwordspacing}{\spaceskip=0pt\relax}
\providecommand{\BIBentryALTinterwordstretchfactor}{4}
\providecommand{\BIBentryALTinterwordspacing}{\spaceskip=\fontdimen2\font plus
\BIBentryALTinterwordstretchfactor\fontdimen3\font minus
  \fontdimen4\font\relax}
\providecommand{\BIBforeignlanguage}[2]{{%
\expandafter\ifx\csname l@#1\endcsname\relax
\typeout{** WARNING: IEEEtran.bst: No hyphenation pattern has been}%
\typeout{** loaded for the language `#1'. Using the pattern for}%
\typeout{** the default language instead.}%
\else
\language=\csname l@#1\endcsname
\fi
#2}}
\providecommand{\BIBdecl}{\relax}
\BIBdecl

\bibitem{RIS_tutorial_2021}
Q.~Wu \emph{et~al.}, ``Intelligent reflecting surface-aided wireless
  communications: A tutorial,'' \emph{IEEE Transactions on Communications},
  vol.~69, no.~5, pp. 3313--3351, 2021.

\bibitem{RIS_THz_2021}
Z.~Chen \emph{et~al.}, ``Intelligent reflecting surface assisted terahertz
  communications toward {6G},'' \emph{IEEE Wireless Communications}, vol.~28,
  no.~6, pp. 110--117, 2021.

\bibitem{RIS_commag_2021}
C.~Pan \emph{et~al.}, ``{Reconfigurable Intelligent Surfaces for {6G} Systems:
  Principles, Applications, and Research Directions},'' \emph{IEEE
  Communications Magazine}, vol.~59, no.~6, pp. 14--20, 2021.

\bibitem{RIS_EE_TWC_2019}
C.~Huang \emph{et~al.}, ``Reconfigurable intelligent surfaces for energy
  efficiency in wireless communication,'' \emph{IEEE Transactions on Wireless
  Communications}, vol.~18, no.~8, pp. 4157--4170, 2019.

\bibitem{distRIS_EE_2022}
Z.~Yang \emph{et~al.}, ``Energy-efficient wireless communications with
  distributed reconfigurable intelligent surfaces,'' \emph{IEEE Transactions on
  Wireless Communications}, vol.~21, no.~1, pp. 665--679, 2022.

\bibitem{DRL_RIS_JSAC_2020}
C.~Huang \emph{et~al.}, ``Reconfigurable intelligent surface assisted multiuser
  {MISO} systems exploiting deep reinforcement learning,'' \emph{IEEE Journal
  on Selected Areas in Communications}, vol.~38, no.~8, pp. 1839--1850, 2020.

\bibitem{RIS_sumrate_2020}
H.~Guo \emph{et~al.}, ``Weighted sum-rate maximization for reconfigurable
  intelligent surface aided wireless networks,'' \emph{IEEE Transactions on
  Wireless Communications}, vol.~19, no.~5, pp. 3064--3076, 2020.

\bibitem{RIS_loc_2021_TWC}
W.~Wang \emph{et~al.}, ``Joint beam training and positioning for intelligent
  reflecting surfaces assisted millimeter wave communications,'' \emph{IEEE
  Transactions on Wireless Communications}, vol.~20, no.~10, pp. 6282--6297,
  2021.

\bibitem{elzanaty2021reconfigurable}
A.~Elzanaty \emph{et~al.}, ``Reconfigurable intelligent surfaces for
  localization: Position and orientation error bounds,'' \emph{IEEE Trans.
  Signal Process.}, vol.~69, pp. 5386--5402, 2021.

\bibitem{wymeersch2020radio}
H.~Wymeersch \emph{et~al.}, ``Radio localization and mapping with
  reconfigurable intelligent surfaces: Challenges, opportunities, and research
  directions,'' \emph{IEEE Vehicular Technology Magazine}, vol.~15, no.~4, pp.
  52--61, 2020.

\bibitem{RIS_SPM_2022}
E.~Björnson \emph{et~al.}, ``Reconfigurable intelligent surfaces: A signal
  processing perspective with wireless applications,'' \emph{IEEE Signal
  Processing Magazine}, vol.~39, no.~2, pp. 135--158, 2022.

\bibitem{LOS_NLOS_NearField_2021}
D.~Dardari \emph{et~al.}, ``{LOS/NLOS} near-field localization with a large
  reconfigurable intelligent surface,'' \emph{IEEE Transactions on Wireless
  Communications}, vol.~21, no.~6, pp. 4282--4294, 2022.

\bibitem{rinchi2022compressive}
O.~Rinchi \emph{et~al.}, ``Compressive near-field localization for multipath
  {RIS}-aided environments,'' \emph{IEEE Communications Letters}, vol.~26,
  no.~6, pp. 1268--1272, 2022.

\bibitem{Cuneyd_WCL_RIS_2022}
C.~Ozturk \emph{et~al.}, ``{RIS}-aided near-field localization under
  phase-dependent amplitude variations,'' \emph{IEEE Transactions on Wireless
  Communications}, accepted for publication, 2023.

\bibitem{Shaban2021}
Z.~Abu-Shaban \emph{et~al.}, ``Near-field localization with a reconfigurable
  intelligent surface acting as lens,'' in \emph{ICC 2021 - IEEE International
  Conference on Communications}, 2021, pp. 1--6.

\bibitem{cuneyd_ICC_RIS_2022}
C.~Öztürk \emph{et~al.}, ``On the impact of hardware impairments on
  {RIS}-aided localization,'' in \emph{ICC 2022 - IEEE International Conference
  on Communications}, 2022, pp. 2846--2851.

\bibitem{rahal2022constrained}
M.~Rahal \emph{et~al.}, ``Constrained {RIS} phase profile optimization and time
  sharing for near-field localization,'' in \emph{2022 IEEE 95th Vehicular
  Technology Conference: (VTC2022-Spring)}, 2022, pp. 1--6.

\bibitem{luan2021phase}
M.~Luan \emph{et~al.}, ``Phase design and near-field target localization for
  {RIS}-assisted regional localization system,'' \emph{IEEE Transactions on
  Vehicular Technology}, vol.~71, no.~2, pp. 1766--1777, 2022.

\bibitem{HRIS_NF_2022}
X.~Zhang \emph{et~al.}, ``Hybrid reconfigurable intelligent surfaces-assisted
  near-field localization,'' \emph{IEEE Communications Letters}, pp. 1--1,
  2022.

\bibitem{mmwave_array_diagnosis_2018}
M.~E. Eltayeb \emph{et~al.}, ``Compressive sensing for millimeter wave antenna
  array diagnosis,'' \emph{IEEE Transactions on Communications}, vol.~66,
  no.~6, pp. 2708--2721, 2018.

\bibitem{ULA_DOA_Sensor_Failure}
B.~Sun \emph{et~al.}, ``Direction-of-arrival estimation under array sensor
  failures with {ULA},'' \emph{IEEE Access}, vol.~8, pp. 26\,445--26\,456,
  2020.

\bibitem{errorAnalysis_RIS_failure_2020}
H.~Taghvaee \emph{et~al.}, ``Error analysis of programmable metasurfaces for
  beam steering,'' \emph{IEEE Journal on Emerging and Selected Topics in
  Circuits and Systems}, vol.~10, no.~1, pp. 62--74, 2020.

\bibitem{RIS_diagnosis_2021}
R.~Sun \emph{et~al.}, ``Diagnosis of intelligent reflecting surface in
  millimeter-wave communication systems,'' \emph{IEEE Transactions on Wireless
  Communications}, vol.~21, no.~6, pp. 3921--3934, 2022.

\bibitem{phaseCalib_RIS_TSP_2022}
J.~Zhang \emph{et~al.}, ``Phase calibration for intelligent reflecting surfaces
  assisted millimeter wave communications,'' \emph{IEEE Transactions on Signal
  Processing}, vol.~70, pp. 1026--1040, 2022.

\bibitem{Doppler_RIS_HWI_2021}
K.~Wang \emph{et~al.}, ``Doppler effect mitigation using reconfigurable
  intelligent surfaces with hardware impairments,'' in \emph{2021 IEEE Globecom
  Workshops (GC Wkshps)}, 2021, pp. 1--6.

\bibitem{Joint_RIS_BS_Diagnosis_2021}
\BIBentryALTinterwordspacing
S.~Ma \emph{et~al.}, ``Joint diagnosis of {RIS} and {BS} for {RIS}-aided
  millimeter-wave system,'' \emph{Electronics}, vol.~10, no.~20, 2021.
  [Online]. Available: \url{https://www.mdpi.com/2079-9292/10/20/2556}
\BIBentrySTDinterwordspacing

\bibitem{Diagnosis_CE_RIS_2020}
B.~Li \emph{et~al.}, ``Joint array diagnosis and channel estimation for
  {RIS}-aided mmwave {MIMO} system,'' \emph{IEEE Access}, vol.~8, pp.
  193\,992--194\,006, 2020.

\bibitem{Fortunati2017}
S.~Fortunati \emph{et~al.}, ``Performance bounds for parameter estimation under
  misspecified models: Fundamental findings and applications,'' \emph{IEEE
  Signal Processing Magazine}, vol.~34, no.~6, pp. 142--157, 2017.

\bibitem{OMP_mmWave_2016}
J.~Lee \emph{et~al.}, ``Channel estimation via orthogonal matching pursuit for
  hybrid {MIMO} systems in millimeter wave communications,'' \emph{IEEE
  Transactions on Communications}, vol.~64, no.~6, pp. 2370--2386, 2016.

\bibitem{mmWave_pos_TWC_2018}
A.~Shahmansoori \emph{et~al.}, ``Position and orientation estimation through
  millimeter-wave {MIMO} in {5G} systems,'' \emph{IEEE Transactions on Wireless
  Communications}, vol.~17, no.~3, pp. 1822--1835, 2018.

\bibitem{grossi2020adaptive}
E.~Grossi \emph{et~al.}, ``Adaptive detection and localization exploiting the
  {IEEE} 802.11ad standard,'' \emph{IEEE Transactions on Wireless
  Communications}, vol.~19, no.~7, pp. 4394--4407, 2020.

\bibitem{RIS_SISO_JSTSP_2022}
K.~Keykhosravi \emph{et~al.}, ``{RIS}-enabled {SISO} localization under user
  mobility and spatial-wideband effects,'' \emph{IEEE Journal of Selected
  Topics in Signal Processing}, pp. 1--1, 2022.

\bibitem{EM_wavefront}
F.~Guidi \emph{et~al.}, ``Radio positioning with {EM} processing of the
  spherical wavefront,'' \emph{IEEE Transactions on Wireless Communications},
  vol.~20, no.~6, pp. 3571--3586, 2021.

\bibitem{nearfieldTrack_TSP_2021}
A.~Guerra \emph{et~al.}, ``Near-field tracking with large antenna arrays:
  Fundamental limits and practical algorithms,'' \emph{IEEE Transactions on
  Signal Processing}, vol.~69, pp. 5723--5738, 2021.

\bibitem{Fresnel_2011}
J.-W. Tao \emph{et~al.}, ``Joint {DOA}, range, and polarization estimation in
  the {Fresnel} region,'' \emph{IEEE Transactions on Aerospace and Electronic
  Systems}, vol.~47, no.~4, pp. 2657--2672, 2011.

\bibitem{Fresnel_2016}
V.~R. Gowda \emph{et~al.}, ``Wireless power transfer in the radiative near
  field,'' \emph{IEEE Antennas and Wireless Propagation Letters}, vol.~15, pp.
  1865--1868, 2016.

\bibitem{CS_AMP_2013}
J.~Ziniel \emph{et~al.}, ``Dynamic compressive sensing of time-varying signals
  via approximate message passing,'' \emph{IEEE Transactions on Signal
  Processing}, vol.~61, no.~21, pp. 5270--5284, 2013.

\bibitem{spike_slab_JMLR_2014}
A.-S. Sheikh \emph{et~al.}, ``A truncated {EM} approach for spike-and-slab
  sparse coding,'' \emph{Journal of Machine Learning Research}, vol.~15, no.~1,
  pp. 2653--2687, 2014.

\bibitem{spike_slab_PAMI_2013}
I.~J. Goodfellow \emph{et~al.}, ``Scaling up spike-and-slab models for
  unsupervised feature learning,'' \emph{IEEE Transactions on Pattern Analysis
  and Machine Intelligence}, vol.~35, no.~8, pp. 1902--1914, 2013.

\bibitem{Fortunati2018Chapter4}
S.~Fortunati \emph{et~al.}, ``{Chapter 4: Parameter bounds under misspecified
  models for adaptive radar detection},'' in \emph{Academic Press Library in
  Signal Processing, Volume 7}, R.~Chellappa \emph{et~al.}, Eds.\hskip 1em plus
  0.5em minus 0.4em\relax Academic Press, 2018, pp. 197--252.

\bibitem{kay1993fundamentals}
S.~M. Kay, \emph{Fundamentals of statistical signal processing: estimation
  theory}.\hskip 1em plus 0.5em minus 0.4em\relax Prentice-Hall, Inc., 1993.

\bibitem{Hybrid_ML_MAP_TSP}
Y.~Noam \emph{et~al.}, ``Notes on the tightness of the hybrid {Cramér–Rao}
  lower bound,'' \emph{IEEE Transactions on Signal Processing}, vol.~57, no.~6,
  pp. 2074--2084, 2009.

\bibitem{wolsey2007mixed}
L.~A. Wolsey, ``Mixed integer programming,'' \emph{Wiley Encyclopedia of
  Computer Science and Engineering}, pp. 1--10, 2007.

\bibitem{Pierre_2012_NPHard}
P.~Bonami \emph{et~al.}, ``Algorithms and software for convex mixed integer
  nonlinear programs,'' in \emph{Mixed Integer Nonlinear Programming}, J.~Lee
  \emph{et~al.}, Eds.\hskip 1em plus 0.5em minus 0.4em\relax New York, NY:
  Springer New York, 2012, pp. 1--39.

\bibitem{liberti2019undecidability}
L.~Liberti, ``Undecidability and hardness in mixed-integer nonlinear
  programming,'' \emph{RAIRO-Operations Research}, vol.~53, no.~1, pp. 81--109,
  2019.

\bibitem{tibshirani2013lasso}
R.~J. Tibshirani, ``The {LASSO} problem and uniqueness,'' \emph{Electronic
  Journal of statistics}, vol.~7, pp. 1456--1490, 2013.

\bibitem{cvx}
M.~Grant \emph{et~al.}, ``{CVX}: Matlab software for disciplined convex
  programming, version 2.1,'' \url{http://cvxr.com/cvx}, Mar. 2014.

\bibitem{beck2009fast}
A.~Beck \emph{et~al.}, ``A fast iterative shrinkage-thresholding algorithm for
  linear inverse problems,'' \emph{SIAM Journal on Imaging Sciences}, vol.~2,
  no.~1, pp. 183--202, Mar. 2009.

\bibitem{mp_93}
S.~Mallat \emph{et~al.}, ``Matching pursuits with time-frequency
  dictionaries,'' \emph{IEEE Transactions on Signal Processing}, vol.~41,
  no.~12, pp. 3397--3415, 1993.

\bibitem{Beck_09_ISTA}
\BIBentryALTinterwordspacing
A.~Beck \emph{et~al.}, ``A fast iterative shrinkage-thresholding algorithm for
  linear inverse problems,'' \emph{SIAM Journal on Imaging Sciences}, vol.~2,
  no.~1, pp. 183--202, 2009. [Online]. Available:
  \url{https://doi.org/10.1137/080716542}
\BIBentrySTDinterwordspacing

\bibitem{TVT_2022_ISTA}
C.~Baquero~Barneto \emph{et~al.}, ``Millimeter-wave mobile sensing and
  environment mapping: Models, algorithms and validation,'' \emph{IEEE
  Transactions on Vehicular Technology}, vol.~71, no.~4, pp. 3900--3916, 2022.

\bibitem{rayleighRician_CommL_2023}
M.~Abbasi~Msleh \emph{et~al.}, ``Ergodic capacity analysis of reconfigurable
  intelligent surface assisted {MIMO} systems over {Rayleigh-Rician}
  channels,'' \emph{IEEE Communications Letters}, vol.~27, no.~1, pp. 75--79,
  2023.

\bibitem{Rician_PhaseOpt_TCOM_2022}
K.~Zhi \emph{et~al.}, ``Power scaling law analysis and phase shift optimization
  of {RIS}-aided massive {MIMO} systems with statistical {CSI},'' \emph{IEEE
  Transactions on Communications}, vol.~70, no.~5, pp. 3558--3574, 2022.

\bibitem{RIS_opt_Rician_2023}
F.~Jiang \emph{et~al.}, ``Two-timescale transmission design and {RIS}
  optimization for integrated localization and communications,'' \emph{IEEE
  Transactions on Wireless Communications}, pp. 1--1, 2023.

\bibitem{IRS_Statistical_2021}
A.~Abrardo \emph{et~al.}, ``Intelligent reflecting surfaces: Sum-rate
  optimization based on statistical position information,'' \emph{IEEE
  Transactions on Communications}, vol.~69, no.~10, pp. 7121--7136, 2021.

\bibitem{peng2021ris}
Z.~Peng \emph{et~al.}, ``{RIS}-aided {D2D} communications relying on
  statistical {CSI} with imperfect hardware,'' \emph{IEEE Communications
  Letters}, vol.~26, no.~2, pp. 473--477, 2021.

\bibitem{rayleighLoc_2021}
A.~Pandey \emph{et~al.}, ``Adaptive mini-batch gradient-ascent-based
  localization for indoor {IoT} networks under {Rayleigh} fading conditions,''
  \emph{IEEE Internet of Things Journal}, vol.~8, no.~13, pp. 10\,665--10\,677,
  2021.

\end{thebibliography}
	
\end{document}